# Extended Computation Tree Logic

Roland Axelsson[1], Matthew Hague[2], Stephan Kreutzer[2], Martin Lange[3], and Markus Latte[1]

[1] Department of Computer Science, Ludwig-Maximilians-Universität Munich,
Email: {roland.axelsson,markus.latte}@ifi.lmu.de
[2] Oxford University Computing Laboratory,
Email: {Matthew.Hague,stephan.kreutzer}@comlab.ox.ac.uk
[3] Department of Computer Science, University of Kassel, Germany,
Email: martin.lange@uni-kassel.de

**Abstract.** We introduce a generic extension of the popular branching-time logic CTL which refines the temporal until and release operators with formal languages. For instance, a language may determine the moments along a path that an until property may be fulfilled. We consider several classes of languages leading to logics with different expressive power and complexity, whose importance is motivated by their use in model checking, synthesis, abstract interpretation, etc.
We show that even with context-free languages on the until operator the logic still allows for polynomial time model-checking despite the significant increase in expressive power. This makes the logic a promising candidate for applications in verification.
In addition, we analyse the complexity of satisfiability and compare the expressive power of these logics to CTL$^*$ and extensions of PDL.

## 1 Introduction

Computation Tree Logic (CTL) is one of the main logical formalisms for program specification and verification. It appeals because of its intuitive syntax and its very reasonable complexities: model checking is P-complete [9] and satisfiability checking is EXPTIME-complete [12]. However, its expressive power is low.

CTL can be embedded into richer formalisms like CTL$^*$ [13] or the modal $\mu$-calculus $\mathcal{L}_\mu$ [20]. This transition comes at a price. For CTL$^*$ the model checking problem increases to PSPACE-complete [28] and satisfiability to 2EXPTIME-complete [14, 31]. Furthermore, CTL$^*$ cannot express regular properties like "something holds after an even number of steps". The modal $\mu$-calculus is capable of doing so, and its complexities compare reasonably to CTL: satisfiability is also EXPTIME-complete, and model checking sits between P and NP$\cap$coNP. However, it is much worse from a pragmatic perspective. For example, its syntax is notoriously unintuitive.

Common to all these (and many other) formalisms is a restriction of their expressive power to at most regular properties. This follows since they can be embedded into (the bisimulation-invariant) fragment of monadic second-order logic on graphs. This restriction yields some nice properties — like the finite model property and decidability — but implies that these logics cannot be used for certain specification purposes.

For example, specifying the correctness of a communication protocol that uses a buffer requires a non-underflow property: an item cannot be removed when the buffer is empty. The specification language must therefore be able to track the buffer's size. If the buffer is unbounded, as is usual in software, this property is non-regular and a regular logic is unsuitable. If the buffer is bounded, the property is regular but depends on the actual buffer capacity, requiring a different formula for each size. This is unnatural for verification purposes. The formulas are also likely to be complex as they essentially have to hard-code numbers up to the buffer length. To express such properties naturally one has to step beyond regularity and consider logics of corresponding expressive power.

A second example is program synthesis, where, instead of verifying a program, one wants to automatically generate a correct program (skeleton) from the specification. This problem is very much linked to satisfiability checking, except, if a model exists, one is created and transformed into a program. This is known as controller synthesis and has been done mainly based on satisfiability checking for $\mathcal{L}_\mu$ [4]. The finite model property restricts the synthesization to finite state programs, i.e. hardware and controllers, etc. In order to automatically synthesize software (e.g. recursive functions) one has to consider non-regular logics.

Finally, a third example occurs when verifying programs with infinite or very large state spaces. A standard technique is to abstract the large state space into a smaller one [10]. This usually results in spurious traces which then have to be excluded in universal path quantification on the small system. If the original system was infinite then the language of spurious traces is typically non-regular and, again, a logic of suitable expressive power is needed to increase precision.

We introduce a generic extension of CTL which provides a specification formalism for such purposes. We refine the usual until operator (and its dual, the release operator) with a formal language defining the moments at which the until property can be fulfilled. This leads to a family of logics parametrised by a class of formal languages. CTL is an ideal base logic because of its wide-spread use in actual verification applications. Since automata easily allow for an unambiguous measure of input size, we present the precise definition of our logics in terms of classes of automata instead of formal languages. However, we do not promote the use of automata in temporal formulas. For pragmatic considerations it may be sensible to allow more intuitive descriptions of formal languages such as Backus-Naur-Form or regular expressions.

As a main result we add context-free languages to the path quantifiers, significantly increasing expressive power, while retaining polynomial time model-checking. Hence, we obtain a good balance between expressiveness — as non-regular properties become expressible — and low model-checking complexity, which makes this logic very promising for applications in verification. We also study model-checking for the new logics against infinite state systems represented by (visibly) pushdown automata, as they arise in software model-checking, and obtain tractability results for these. For satisfiability testing, equipping the path quantifiers with visibly pushdown languages retains decidability. However, the complexity increases from EXPTIME for CTL to 3EXPTIME for this new logic.

The paper is organised as follows. We formally introduce the logics and give an example demonstrating their expressive power in Sect. 2. Sect. 3 discusses related formalisms. Sect. 4 presents results on the expressive power of these logics, and Sect. 5 and 6 contain results on the complexities of satisfiability and model checking. Finally, Sect. 7 concludes with remarks on further work.

## 2 Extended Computation Tree Logic

Let $\mathcal{P} = \{p, q, \ldots\}$ be a countably infinite set of *propositions* and $\Sigma$ be a finite set of *action names*. A *labeled transition system* (LTS) is a $\mathcal{T} = (\mathcal{S}, \rightarrow, \ell)$, where $\mathcal{S}$ is a set of states, $\rightarrow\ \subseteq \mathcal{S} \times \Sigma \times \mathcal{S}$ and $\ell : \mathcal{S} \rightarrow 2^{\mathcal{P}}$. We usually write $s \xrightarrow{a} t$ instead of $(s, a, t) \in \rightarrow$. A *path* is a maximal sequence of alternating states and actions $\pi = s_0, a_1, s_1, a_2, s_2, \ldots$, s.t. $s_i \xrightarrow{a_{i+1}} s_{i+1}$ for all $i \in \mathbb{N}$. We also write a path as $s_0 \xrightarrow{a_1} s_1 \xrightarrow{a_2} s_2 \ldots$ Maximality means that the path is either infinite or it ends in a state $s_n$ s.t. there are no $a \in \Sigma$ and $t \in \mathcal{S}$ with $s_n \xrightarrow{a} t$. In the latter case, the domain $dom(\pi)$ of $\pi$ is $\{0, \ldots, n\}$. And otherwise $dom(\pi):=\mathbb{N}$.

We focus on automata classes between deterministic finite automata (DFA) and nondeterministic pushdown automata (PDA), with the classes of nondeterministic finite automata (NFA), (non-)deterministic visibly pushdown automata (DVPA/VPA) [2] and deterministic pushdown automata (DPDA) in between. Beyond PDA one is often faced with undecidability. Note that some of these automata classes define the same class of languages. However, translations from nondeterministic to deterministic automata usually involve an exponential blow-up. For complexity estimations it is therefore advisable to consider such classes separately.

We call a class $\mathfrak{A}$ of automata *reasonable* if it contains automata recognising $\Sigma$ and $\Sigma^*$ and is closed under equivalences, i.e. if $\mathcal{A} \in \mathfrak{A}$ and $L(\mathcal{A}) = L(\mathcal{B})$ and $\mathcal{B}$ is of the same type then $\mathcal{B} \in \mathfrak{A}$. $L(\mathcal{A})$ denotes the language accepted by $\mathcal{A}$.

Let $\mathfrak{A}, \mathfrak{B}$ be two reasonable classes of finite-word automata over the alphabet $\Sigma$. Formulas of *Extended Computation Tree Logic over* $\mathfrak{A}$ *and* $\mathfrak{B}$ (CTL[$\mathfrak{A}$,$\mathfrak{B}$]) are given by the following grammar, where $\mathcal{A} \in \mathfrak{A}$, $\mathcal{B} \in \mathfrak{B}$ and $q \in \mathcal{P}$.

$$\varphi \ ::= \ q \mid \varphi \vee \varphi \mid \neg\varphi \mid \mathtt{E}(\varphi \mathtt{U}^{\mathcal{A}} \varphi) \mid \mathtt{E}(\varphi \mathtt{R}^{\mathcal{B}} \varphi)$$

Formulas are interpreted over states of a transition system $\mathcal{T} = (\mathcal{S}, \rightarrow, \ell)$ in the following way.

- $\mathcal{T}, s \models q$ iff $q \in \ell(s)$
- $\mathcal{T}, s \models \varphi \vee \psi$ iff $\mathcal{T}, s \models \varphi$ or $\mathcal{T}, s \models \psi$ and $\mathcal{T}, s \models \neg\varphi$ iff $\mathcal{T}, s \not\models \varphi$
- $\mathcal{T}, s \models \mathtt{E}(\varphi \mathtt{U}^{\mathcal{A}} \psi)$ iff there exists a path $\pi = s_0, a_1, s_1, \ldots$ with $s_0 = s$ and $\exists n \in dom(\pi)$ s.t. $a_1 \ldots a_n \in L(\mathcal{A})$ and $\mathcal{T}, s_n \models \psi$ and $\forall i < n : \mathcal{T}, s_i \models \varphi$.
- $\mathcal{T}, s \models \mathtt{E}(\varphi \mathtt{R}^{\mathcal{A}} \psi)$ iff there exists a path $\pi = s_0, a_1, s_1, \ldots$ with $s_0 = s$ and for all $n \in dom(\pi)$: $a_1 \ldots a_n \notin L(\mathcal{A})$ or $\mathcal{T}, s_n \models \psi$ or $\exists i < n$ s.th. $\mathcal{T}, s_i \models \varphi$.

As usual, further syntactical constructs, like other boolean operators, are introduced as abbreviations. Similarly, we define $\mathtt{A}(\varphi \mathtt{U}^{\mathcal{A}} \psi) := \neg\mathtt{E}(\neg\varphi \mathtt{R}^{\mathcal{A}} \neg\psi)$, $\mathtt{A}(\varphi \mathtt{R}^{\mathcal{A}} \psi) := \neg\mathtt{E}(\neg\varphi \mathtt{U}^{\mathcal{A}} \neg\psi)$, as well as $Q\mathtt{F}^{\mathcal{A}}\varphi := Q(\mathtt{tt}\mathtt{U}^{\mathcal{A}}\varphi)$, $Q\mathtt{G}^{\mathcal{A}}\varphi := Q(\mathtt{ff}\mathtt{R}^{\mathcal{A}}\varphi)$ for $Q \in \{\mathtt{E}, \mathtt{A}\}$. For presentation, we also use languages $L$ instead of automata in the temporal operators. For instance, $\mathtt{EG}^{L}\varphi$ is $\mathtt{EG}^{\mathcal{A}}\varphi$ for some $\mathcal{A}$ with $L(\mathcal{A}) = L$. This also allows us to easily define the original CTL operators: $Q\mathtt{X}\varphi := Q\mathtt{F}^{\Sigma}\varphi$, $Q(\varphi \mathtt{U} \psi) := Q(\varphi \mathtt{U}^{\Sigma^*} \psi)$, $Q(\varphi \mathtt{R} \psi) := Q(\varphi \mathtt{R}^{\Sigma^*} \psi)$, etc. The size of a formula $\varphi$ is the number of its unique subformulas plus the sum of the sizes of all automata in $\varphi$, with the usual measure of size of an automaton.

The distinction between $\mathfrak{A}$ and $\mathfrak{B}$ is motivated by the complexity analysis. For instance, when model checking $\mathtt{E}(\varphi \mathtt{U}^{\mathcal{A}} \psi)$ the existential quantifications over system paths and runs of $\mathcal{A}$ commute and we can step-wise guess a path and an accepting run. On the other hand, when checking $\mathtt{E}(\varphi \mathtt{R}^{\mathcal{A}} \psi)$ the existential quantification on paths and universal quantification on runs (by $\mathtt{R}$ – "on all prefixes ...") does not commute unless we determinise $\mathcal{A}$, which is not always possible or may lead to exponential costs.

However, $\mathfrak{A}$ and $\mathfrak{B}$ can also be the same and in this case we denote the logic by CTL[$\mathfrak{A}$]. Equally, by EF[$\mathfrak{A}$], resp. EG[$\mathfrak{B}$] we denote the fragments of CTL[$\mathfrak{A}$,$\mathfrak{B}$] built from atomic propositions, boolean operators and the temporal operators $\mathtt{EF}^{\mathcal{A}}\varphi$, resp. $\mathtt{EG}^{\mathcal{B}}\varphi$ only. Since the expressive power of the logic only depends on its class of *languages* rather than *automata*, we will write CTL[REG], CTL[VPL], CTL[CFL], etc. to denote the logic over regular, visibly pushdown, and context-free languages, represented by any type of automaton.

*Example.* We close this section with a CTL[VPL] example demonstrating the buffer-underflow property discussed in the introduction. Consider a concurrent producer/consumer scenario over a shared buffer. If the buffer is empty, the consumer process requests a new resource and halts until the producer delivers a new one. Any parallel execution of these processes should obey a non-underflow property (NBU): at any moment, the number of produce actions is sufficient for the number of consumes.

If the buffer is realised in software it is reasonable to assume that it is unbounded, and thus, non-regular. Let $\Sigma = \{p, c, r\}$, where $p$ stands for *production* of a buffer object, $c$ for *consume* and $r$ for *request*. The NBU property is given by the VPL $L = \{w \in \Sigma^* \mid \ |w|_c = |w|_p \text{ and } |v|_c \leq |v|_p \text{ for all } v \preceq w\}$, where $\preceq$ denotes the prefix relation. We express the requirements in CTL[VPL].

1. $\mathtt{AGEX}^{p}\mathtt{tt}$ : "at any time it is possible to produce an object"
2. $\mathtt{AG}^{L}(\mathtt{AX}^{c}\mathtt{ff} \wedge \mathtt{EX}^{r}\mathtt{tt})$: "whenever the buffer is empty, it is impossible to consume and possible to request"
3. $\mathtt{AG}^{\overline{L}}(\mathtt{EX}^{c}\mathtt{tt} \wedge \mathtt{AX}^{r}\mathtt{ff})$: "whenever the buffer is non-empty it is possible to consume and impossible to request"
4. $\mathtt{EFEG}^{\overline{c^*}}\mathtt{ff}$: "at some point there is a consume-only path"

Combining the first three properties yields a specification of the scenario described above and states that a *request* can only be made if the buffer is empty. For the third properly, recall that VPL are closed under complement [2]. Every satisfying model gives a raw implementation of the main characteristics of the system. Note that if it is always possible to *produce* and possible to *consume* iff the buffer is not empty, then a straight-forward model with self-loops $p, c$ and $r$ does not satisfy the specification. Instead, we require a model with infinitely many different $p$ transitions. If we strengthen the specification by adding the fourth formula, it becomes unsatisfiable.

## 3 Related Formalisms

Several suggestions to integrate formal languages into temporal logics have been made so far. The goal is usually to extend the expressive power of a logic whilst retaining its intuitive syntax. A classic example is Propositional Dynamic Logic (PDL) [16] which extends Modal Logic with regular expressions. There are similar extensions of LTL [32, 18, 21] and of CTL [5, 7, 25] refining the temporal operators with regular languages in some form. The need for extensions beyond the use of pure temporal operators is also witnessed by the industry-standard *Property Specification Language* (PSL) [1] and its predecessor ForSpec [3]. However, both logics do not reach beyond regular properties, whereas we are mostly interested in non-regular properties. Furthermore, ForSpec is a linear-time formalism, whereas we are concerned with branching-time. PSL does contain branching-time operators but they have been introduced for backwards-compatibility only, and nowadays, PSL is also mainly understood to be linear-time.

While much effort has been put into regular extensions of standard temporal logics, little is known about extensions using richer classes of formal languages. We are only aware of extensions of PDL by context-free languages [17] or visibly pushdown languages [23].

The main yardstick for measuring the expressive power of CTL[$\mathfrak{A}$,$\mathfrak{B}$] will be PDL and one of its variants, namely PDL with the $\Delta$-construct and tests, $\Delta$PDL$^?$[REG], [16, 29]. This was in detail only considered for regular, visibly pushdown and context-free languages so far, but it can of course be straightforwardly extended to arbitrary classes of formal languages represented by automata. It is, however, necessary to define automata over a potentially unbounded alphabet arising from the interleaving of letters with tests in a word.

We briefly recall its syntax and semantics here as it will be needed below.

**Semantics of PDL with the $\Delta$-operator and tests.** Formulas $\mathsf{Form}$ and programs $\mathsf{Prog}$ of $\Delta$PDL$^?$[$\mathfrak{A}$] for some $\mathfrak{A}$ over an alphabet $\Sigma$ are the least sets satisfying the following.

1. $\mathcal{P} \subseteq \mathsf{Form}$.
2. If $\varphi, \psi \in \mathsf{Form}$ then $\varphi \vee \psi \in \mathsf{Form}, \neg\varphi \in \mathsf{Form}$.
3. If $\varphi \in \mathsf{Form}$, $\mathcal{A} \in \mathsf{Prog}$ then $\langle\mathcal{A}\rangle\varphi \in \mathsf{Form}$.
4. $\mathfrak{A} \subseteq \mathsf{Prog}$.
5. For every $\mathfrak{A}$-automaton $\mathcal{A}$ over $\Sigma \cup \{\varphi? \mid \varphi \in \mathsf{Form}\}$ we have $\mathcal{A} \in \mathsf{Prog}$.
6. If $\mathcal{A} \in \mathsf{Prog}$ and $\mathcal{A}'$ results from $\mathcal{A}$ by equipping it with a Büchi condition on states, then $\Delta\mathcal{A}' \in \mathsf{Form}$.

$\Delta$PDL$^?[\mathfrak{A}]$ consists of all elements of $\mathsf{Form}$ which are constructed in this way. The fragment PDL$[\mathfrak{A}]$ is obtained by removing clauses (5) and (6). The semantics is again defined over states of transition systems. The clauses for atomic propositions and the boolean operators are as usual. For the other constructs, we use the fact that programs and formulas are defined inductively. For a $\mathcal{T} = (\mathcal{S}, \rightarrow, \ell)$ with edge labels in $\Sigma'$ and a finite subset $\Phi \subset \mathsf{Form}$ of formulas let $\mathcal{T}^{\Phi}$ result from $\mathcal{T}$ by adding, for every $s \in \mathcal{S}$ and every $\varphi \in \Phi$, a transition $s \xrightarrow{\varphi?} s$ if $\mathcal{T}, s \models \varphi$. For a formula $\varphi$ let $?(\varphi)$ be the set of all tests $\psi?$ occurring in $\varphi$ syntactically.

$$\begin{aligned}
\mathcal{T}, s \models \langle \mathcal{A} \rangle \varphi \text{ iff } & \exists \pi = s_0 \xrightarrow{a_1} s_1 \xrightarrow{a_2} \ldots \xrightarrow{a_n} s_n \text{ in} \\
& \mathcal{T}^{?(\langle \mathcal{A} \rangle \varphi)} \text{ with } s_0 = s \text{ s.t.} \\
& \text{(i) } a_1 \ldots a_n \in L(\mathcal{A}), \text{ and} \\
& \text{(ii) } \mathcal{T}, s_n \models \varphi. \\
\mathcal{T}, s \models \Delta \mathcal{A} \text{ iff } & \exists \pi = s_0 \xrightarrow{a_1} s_1 \xrightarrow{a_2} \ldots \text{ in } \mathcal{T}^{?(\langle \mathcal{A} \rangle \varphi)} \\
& \text{with } s_0 = s \text{ and } a_1 a_2 \ldots \in L(\mathcal{A}).
\end{aligned}$$

In the following we will use some of the known results about $\Delta$PDL$^?[\mathfrak{A}]$ for some classes $\mathfrak{A}$ of automata.

**Theorem 3.1.** *Satisfiability testing for*

1. $\Delta$PDL$^?$[Reg] *is in EXPTIME [15].*
2. $\Delta$PDL$^?$[VPL] *is in 2EXPTIME [23].*
3. CTL *is EXPTIME-hard [12].*
4. PDL[DVPA] *is 2EXPTIME-hard [23].*

There are also temporal logics which obtain higher expressive power through other means. These are usually extensions of $\mathcal{L}_\mu$ like the Modal Iteration Calculus [11] which uses inflationary fixpoint constructs or Higher-Order Fixpoint Logic [33] which uses higher-order predicate transformers. While most regular extensions of standard temporal logics like CTL and LTL can easily be embedded into $\mathcal{L}_\mu$, little is known about the relationship between richer extensions of these logics.

## 4 Expressivity and Model Theory

We write $\mathcal{L} \leq_f \mathcal{L}'$ with $f \in \{\mathsf{lin}, \mathsf{exp}\}$ to state that for every formula $\varphi \in \mathcal{L}$ there is an equivalent $\psi \in \mathcal{L}'$ with at most a linear or exponential (respectively) blow up in size. We use $\lneq_f$ to denote that such a translation exists, but there are formulas of $\mathcal{L}'$ which are not equivalent to any formla in $\mathcal{L}$. Also, we write $\mathcal{L} \equiv_f \mathcal{L}'$ if $\mathcal{L} \leq_f \mathcal{L}'$ and $\mathcal{L}' \leq_f \mathcal{L}$. We will drop the index if a potential blow-up is of no concern.

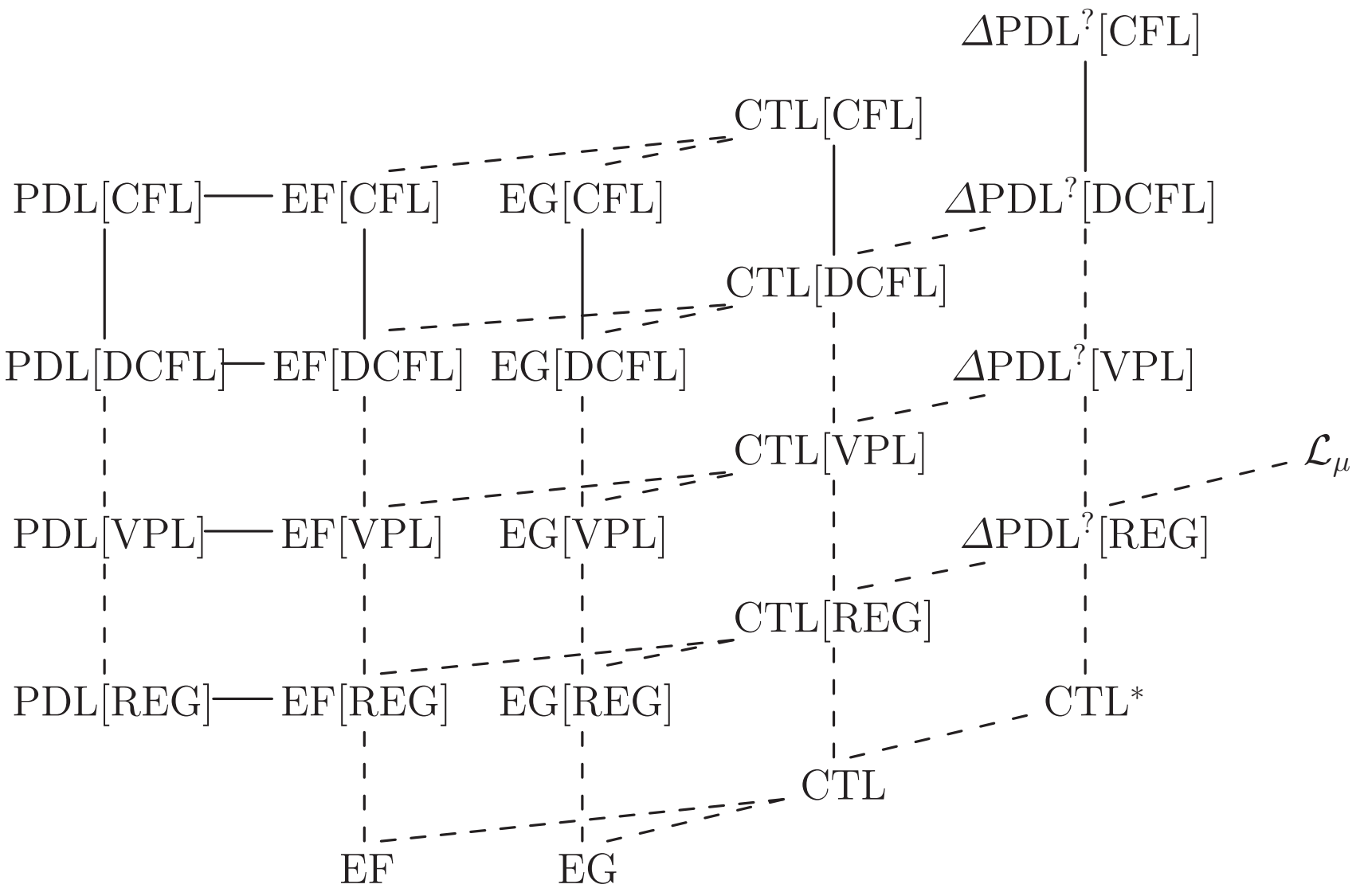

**Fig. 1.** The expressive power of Extended Computation Tree Logic.

A detailed picture of the expressivity results regarding the most important CTL[$\mathfrak{A}$] logics is given in Fig. 1. A (dashed) line moving upwards indicates (strict) inclusion w.r.t. expressive power. A horizontal continuous line states expressive equivalence. It is currently unknown whether or not CTL[CFL] is embeddable into $\Delta$PDL$^?$[CFL]. If such an embedding exists, it has to be strict. Note that CFL does not admit deterministic automata. Hence, Thm. 4.2 (3) (given below) is not applicable in this case.

The following proposition collects some simple observations.

**Proposition 4.1.** *1. For all* $\mathfrak{A}, \mathfrak{B}$*:* CTL $\lneq_{\mathsf{lin}}$ CTL[$\mathfrak{A}$,$\mathfrak{B}$].
*2. For all* $\mathfrak{A}, \mathfrak{A}', \mathfrak{B}, \mathfrak{B}'$*: if* $\mathfrak{A} \leq \mathfrak{A}'$ *and* $\mathfrak{B} \leq \mathfrak{B}'$ *then* CTL[$\mathfrak{A}$,$\mathfrak{B}$] $\leq$ CTL[$\mathfrak{A}'$,$\mathfrak{B}'$].

*Proof.* 1. CTL can be embedded into CTL[$\mathfrak{A}$,$\mathfrak{B}$] because $\mathfrak{A}$ is assumed to contain automata for the languages $\Sigma$ and $\Sigma^*$. It is then easy to translate CTL formulas inductively. Atomic propositions and boolean operators are translated straight-forwardly. The temporal operators are translated using $\mathtt{E}(\varphi \mathtt{U} \psi) \equiv \mathtt{E}(\varphi \mathtt{U}^{\Sigma^*} \psi)$, $\mathtt{E}(\varphi \mathtt{R} \psi) \equiv \mathtt{E}(\varphi \mathtt{R}^{\Sigma^*} \psi)$, and $\mathtt{EX}\varphi \equiv \mathtt{EF}^{\Sigma}\varphi$. Note that $\Sigma$ and $\Sigma^*$ can be defined by NFA of constant size. Hence, the blow-up is linear. Strictness of the inclusion is a simple consequence of the fact that CTL is usually interpreted over transition systems with no transition labels. Hence, $\mathtt{EX}^a\mathtt{tt} := \mathtt{EF}^{\{a\}}\mathtt{tt}$ says that a state has an $a$-successor and is therefore not expressible in CTL. However, it is also known that properties like $\mathtt{AG}^{(aa)^*} q$ – proposition $q$ holds on all even moments – are not expressible in CTL.
2. Clearly, with more automata as arguments of the temporal operators one cannot decrease the expressive power. □

CTL[$\mathfrak{A}$] is properly situated between PDL[$\mathfrak{A}$] and $\Delta$PDL$^?$[$\mathfrak{A}$]. In fact, PDL[$\mathfrak{A}$] is just a syntactic variation of the EF[$\mathfrak{A}$] fragment. The upper bound imposed by $\Delta$PDL$^?$[$\mathfrak{A}$], however, only holds for certain classes $\mathfrak{A}$.

**Theorem 4.2.** *1. For all $\mathfrak{A}$:* PDL[$\mathfrak{A}$] $\equiv_{\mathsf{lin}}$ EF[$\mathfrak{A}$].

*2. For all $\mathfrak{A}, \mathfrak{B}$:* EF[$\mathfrak{A}$] $\lneq_{\mathsf{lin}}$ CTL[$\mathfrak{A}$,$\mathfrak{B}$].

*3. For all $\mathfrak{A}, \mathfrak{B}$: if $\mathfrak{B}$ is a class of deterministic automata then* CTL[$\mathfrak{A}$,$\mathfrak{B}$] $\leq_{\mathsf{lin}}$ $\Delta$PDL$^?$[$\mathfrak{A} \cup \mathfrak{B}$].

*Proof.* 1. Both embeddings of PDL[$\mathfrak{A}$] into EF[$\mathfrak{A}$] and back are done by a simple induction on the structure of formulas. Note that the PDL[$\mathfrak{A}$] formulas do not contain tests nor the $\Delta$ operator. Both inductions are easily carried out using the equivalence $\langle\mathcal{A}\rangle\varphi \equiv \mathtt{EF}^{\mathcal{A}}\varphi$.

2. Clearly, EF[$\mathfrak{A}$] is a fragment of CTL[$\mathfrak{A}$,$\mathfrak{B}$] for any $\mathfrak{B}$. In order to show strictness of the inclusion we extend – in a straight-forward manner – the proof of the corresponding result for the logics EF and CTL [12]. First we define the quotient of a transition system $\mathcal{T} = (\mathcal{S}, \rightarrow, \ell)$ under a set of formulas $\Phi \subseteq$ EF[$\mathfrak{A}$]. It is $\mathcal{T}/\Phi = (\mathcal{S}/\Phi, \rightarrow, \ell/\Phi)$ with
   - $\mathcal{S}/\Phi = \{[s] \mid s \in \mathcal{S}\}$ where $[s] = \{t \in \mathcal{S} \mid s \sim_\Phi t\}$ and $s \sim_\Phi t$ iff $\forall\varphi \in \Phi : \mathcal{T}, s \models \varphi$ iff $\mathcal{T}, t \models \varphi\}$,
   - $[s] \xrightarrow{a} [t]$ iff $\exists s', t'$ with $s' \sim_\Phi s$, $t' \sim_\Phi t$, and $s' \xrightarrow{a} t'$,
   - $\ell/\Phi([s]) = \ell(s) \cap \Phi$.

   It is then easy to show that for all $\mathcal{T}$ with states $s$ and all $\varphi \in$ EF[$\mathfrak{A}$] we have $\mathcal{T}, s \models \varphi$ iff $\mathcal{T}/Sub(\varphi), [s] \models \varphi$. Furthermore, $|\mathcal{S}/Sub(\varphi)| \leq 2^{|Sub(\varphi)|}$. On the other hand, consider the transition system $\mathcal{T} = (\mathbb{N}, \rightarrow, \ell)$ with $i \rightarrow j$ iff $j = i - 1$, and $\ell(i) = \emptyset$ for all $i \in \mathbb{N}$. Clearly, we have $\mathcal{T}, i \models \mathtt{AFAXff}$ for all $i \in \mathbb{N}$. However, suppose there was an EF[$\mathfrak{A}$] formula $\varphi_0$ equivalent to $\mathtt{AFAXff}$. Then $\mathcal{T}/Sub(\varphi_0), [i] \not\models \varphi_0$ for some $i > 2^{|Sub(\varphi)|}$ because this quotient must contain a cycle without the state [0]. But this contradicts the fact that EF[$\mathfrak{A}$]-definable properties are invariant under these quotients.

3. We focus on finite state automaton only. However, the proof can be extended to the two kinds of pushdown automata considered in the report—every subsequent replacement can be extended to push, pop and internal operations. Tests are internal operations, anyway. The proposed translation of the $\mathtt{ER}$-formulas relies on an translation of the possibly larger formula $\mathtt{AXff} \equiv \neg\mathtt{E}(\mathtt{tt}\mathtt{U}^{\Sigma}\mathtt{tt})$. As latter does not involve any $\mathtt{ER}$-formula we may assume an appropriate induction principle. The translations of proposition and boolean operation are straight forward. Given a CTL-formula $\mathtt{E}(\psi_1\mathtt{U}^{\mathcal{A}}\psi_2)$, we construct an automaton $\mathcal{A}'$ by modifying $\mathcal{A}$ as follows.

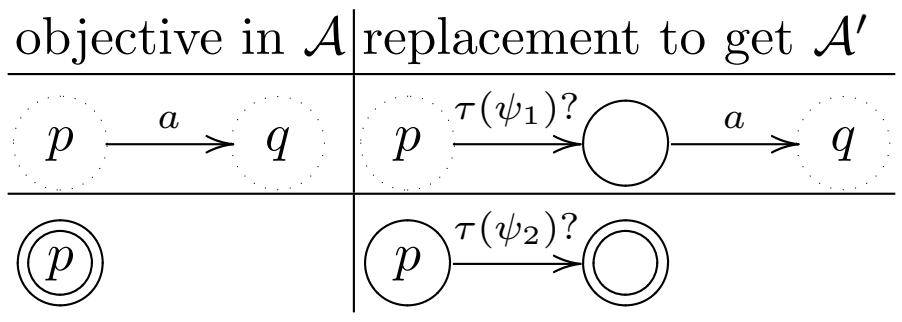

Each dotted circle matches either a final or non-final state. The function $\tau$ refers to the translation for those formulas for which the induction hypothesis is applicable. Obviously $\mathcal{T}, s \models \mathtt{E}(\psi_1 \mathtt{U}^{\mathcal{A}} \psi_2)$ iff $\mathcal{T}, s \models \langle \mathcal{A}' \rangle \mathtt{tt}$.
And as for a formula $\varphi{:=}\mathtt{E}(\psi_1 \mathtt{R}^{\mathcal{A}} \psi_2)$, the automaton $\mathcal{A}$ is assumed to be complete, and is turned into a safety $\omega$-automaton $\mathcal{A}'$ as follows. The translation of $\varphi$ is $\Delta\mathcal{A}'$.

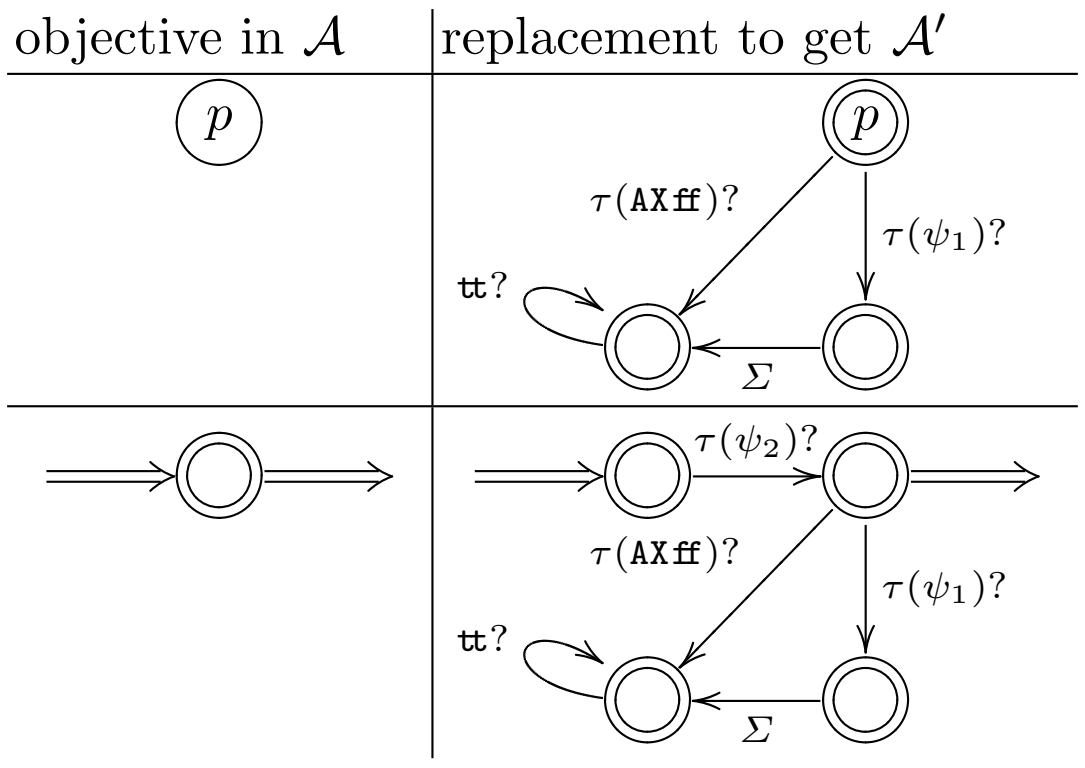

The double arrows indicate either in- or outgoing edges. For the later discussions we wish $\mathcal{A}'$ to be deterministic. Therefore, the edge $\tau(\psi_1)?$ is omitted iff $\tau(\mathtt{AXff}) = \tau(\psi_1)$. Note that in this case, the $\Sigma$-transition is not eligible anyway as $\tau(\psi_1)?$ reports a dead-end state in the LTS.
Let $\pi = s_0, a_1, s_1, \ldots$ be a path witnessing $\mathcal{T}, s_0 \models \mathtt{E}(\psi_1 \mathtt{R}^{\mathcal{A}} \psi_2)$. As $\mathcal{A}$ is deterministic, the run of $\mathcal{A}$ on a prefix $\pi'$ of $\pi$ is a prefix of the run on $\pi$. Thus the witnessing path can be turned into a run in $\mathcal{A}'$: As long as $\psi_1$ does not hold we follow the trace of $\mathcal{A}$. In particular, if a final state in $\mathcal{A}$ is reached then the respective edge $\tau(\psi_2)?$ can be passed. Now, if the current state has no children then $\mathcal{A}'$ can follow the edges $\tau(\mathtt{AXff})?$, and for ever $\mathtt{tt}?$ . And if $\psi_1$ holds in the current state of the LTS the proof obligation vanishes for the next state onwards. Hence $\mathcal{A}'$ can take the way $\tau(\psi_1)?$.
Conversely, let $\pi = s_0, a_1, s_1, \ldots$ be a path witnessing $\mathcal{T}, s_0 \models \Delta\mathcal{A}'$. The run on this path has a prefix—maybe the whole run—which corresponds to a run of $\mathcal{A}$ on $\pi$ where $\psi_2$ is ensured every time $\mathcal{A}$ recognizing the present word, that is, the states in the lower line of replacement are not taken. If the prefix is infinite the run is an infinite witness for $\varphi$. Otherwise, the suffix has the shape $(\tau(\mathtt{AXtt})?)\ (\mathtt{tt}?)^*$ or $(\tau(\psi_1)?)\ \Sigma\ (\mathtt{tt}?)^*$. Both alternatives presents a finite witness of $\varphi$. In particular, the last state of the first witness has no successors.
Finally, the transition is only linearly increasing.

□

If for some classes $\mathfrak{A}$,$\mathfrak{B}$ the inclusion in Part 3 holds, then it must be strict. This is because fairness is not expressible in CTL[$\mathfrak{A}$] regardless of what $\mathfrak{A}$ is, as demonstrated by the following lemma.

**Lemma 4.3.** *The* CTL$^*$*-formula* $\mathtt{EGF}q$ *expressing fairness is not equivalent to any* CTL[$\mathfrak{A}$, $\mathfrak{B}$] *formula, for any* $\mathfrak{A}$, $\mathfrak{B}$.

*Proof.* To simplify notation we denote by $\varphi_{fair}$ the CTL$^*$-formula $\mathtt{EGF}q$ which can equivalently be defined by the $\Delta$PDL$^?$[REG] formula $\Delta\mathcal{A}_{\mathsf{fair}}$.

For $n, k \geq 0$ we define a pair of transition systems $\mathcal{T}_{n,k}$ and $\mathcal{S}_{n,k}$ as follows. For $n = 0$, we define $\mathcal{T}_{0,k}$ and $\mathcal{S}_{0,k}$ as

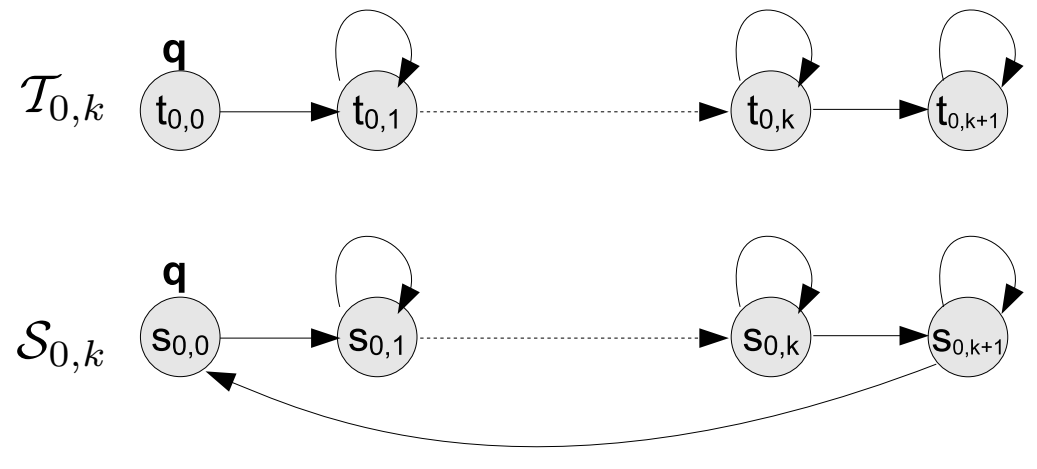

For $n > 0$, we define $\mathcal{T}_{n,k}$ and $\mathcal{S}_{n,k}$ as .

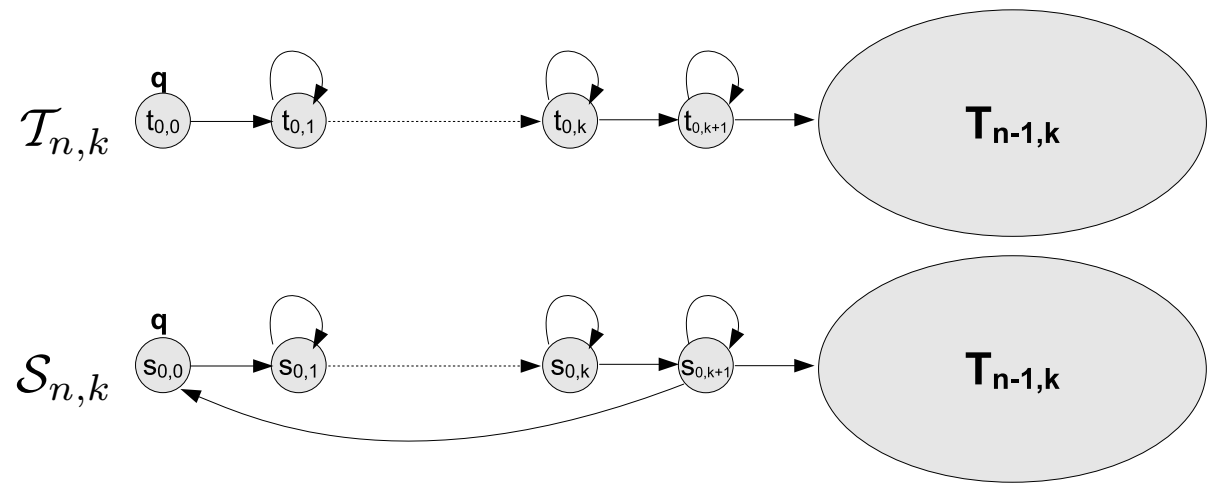

Formally, $\mathcal{T}_{n,k} := (T, \rightarrow, l)$ where $T := \{t_{l,m} : 0 \leq l \leq n, 0 \leq m \leq k+1\}$ and

- $t_{l,m} \rightarrow t_{l,m}$ for all $l$ and $0 < m \leq k+1$,
- $t_{l,m} \rightarrow t_{l,m+1}$ for all $0 \leq l \leq n$ and $0 \leq m < k+1$ and
- $t_{l,k+1} \rightarrow t_{l-1,0}$ for $1 \leq l \leq n$.

Finally, $l(t_{l,0}) = \{q\}$ for all $l$ and $l(x) := \emptyset$ for all other states $x$. $\mathcal{S}_{n,k} := (S, \rightarrow, l)$ is defined analogously over the set $S := \{s_{l,m} : 0 \leq l \leq n, 0 \leq m \leq k+1\}$ of states but in addition there is a transition $s_{n,k+1} \rightarrow s_{n,0}$. Note that there are no actions at the transitions so that automata in CTL[$\mathfrak{A}$,$\mathfrak{B}$]-formulas run over a unary alphabet.

We first establish a simple property of the transition systems defined above which follows immediately from the fact that if $r < \min\{n, n'\}$ then $\mathcal{S}_{n',k}, s_{r,d} \sim \mathcal{T}_{n,k}, t_{r,d}$ for all $d$.

*Claim 1.* If $\chi \in$ CTL[$\mathfrak{A}$,$\mathfrak{B}$] and $r < \min\{n, n'\}$ then for all $0 \leq d \leq k+1$

$$\mathcal{T}_{n,k}, t_{r,d} \models \chi \iff \mathcal{S}_{n,k}, s_{r,d} \models \chi.$$

Let $\mathcal{M} := \{(\mathcal{T}, s) : \mathcal{T}, s \models \varphi_{fair}\}$. Clearly, $\mathcal{S}_{n,k}, s_{n,0} \in \mathcal{M}$ whereas $\mathcal{T}_{n,k}, s_{n,0} \notin \mathcal{M}$, for all $n, k$. We show next that for every $\varphi \in$ CTL[$\mathfrak{A}$,$\mathfrak{B}$] there are $n, k$ such that $\mathcal{T}_{n,k}, t_{n,0} \models \varphi$ if, and only if, $\mathcal{S}_{n,k}, s_{n,0} \models \varphi$. This clearly implies that there is no CTL[$\mathfrak{A}$,$\mathfrak{B}$]-formula defining $\mathcal{M}$ and therefore proves the theorem.

Towards this aim, we define the following two complexity measures for formulas CTL[$\mathfrak{A}$,$\mathfrak{B}$]. W.l.o.g. we assume that $\mathcal{L}(\mathcal{A}) \neq \emptyset$ and $\varepsilon \notin \mathcal{L}(\mathcal{A})$ for all automata $\mathcal{A}$ occurring in $\varphi$, where $\varepsilon$ denotes the empty word.

– The *temporal depth* $td(\varphi)$ is defined in the usual way as the maximal nesting depth of temporal operators in $\varphi$.
– Let $\mathcal{A}$ be an automaton occurring in $\varphi$. If $\mathcal{L}(\mathcal{A})$ is finite we define $ad(\mathcal{A}) := \max\{|w| : w \in \mathcal{L}(\mathcal{A})\}$. Otherwise, $ad(\mathcal{A}) := \min\{|w| : w \in \mathcal{L}(\mathcal{A})\}$.
The *automata depth* $ad(\varphi)$ is defined as $\max\{ad(\mathcal{A}) : \mathcal{A}$ occurs in $\varphi\}$.

*Claim 2.* If $n, n', k \geq 0$ are such that $k > ad(\varphi)$ and $n, n' > td(\varphi)$ then for all $0 \leq m \leq k+1$

$$\mathcal{T}_{n,k}, t_{n,m} \models \varphi \text{ if, and only if, } \mathcal{S}_{n',k}, s_{n',m} \models \varphi.$$

The claim is proved by induction on the structure of formulas $\varphi \in \mathrm{CTL}[\mathfrak{A},\mathfrak{B}]$. This is obvious for atomic formulas and follows immediately from the induction hypothesis for Boolean combinations $\neg\varphi, \varphi_1 \vee \varphi_2, \varphi_1 \wedge \varphi_2$.

**Suppose** $\vartheta := \mathtt{E}(\varphi \mathtt{R}^{\mathcal{A}} \psi)$**.** Let $n, n', k \geq 0$ be such that $k > ad(\varphi)$ and $n, n' > td(\vartheta)$. Suppose $\mathcal{S}_{n',k}, s_{n',0} \models \vartheta$ and let $\pi' := v_0 v_1 \ldots \models \varphi \mathtt{R}^{\mathcal{A}} \psi$, with $v_0 = s_{n',0}$, be a path witnessing this. Let $w$ be a shortest word in $\mathcal{L}(\mathcal{A})$. By construction, $0 < |w| \leq k$. Let $s_{n',d} := v_{|w|}$ for some $0 \leq d \leq k$. Suppose first that $\mathcal{S}_{n',k}, s_{n',d} \models \psi$. By induction hypothesis, $\mathcal{T}_{n,k}, t_{n,d} \models \psi$ and therefore the path $\pi := t_{n,0} \cdot t_{n,1} \cdots t_{n,d} (t_{n,d})^{\omega}$ witnesses that $\mathcal{T}_{n,k} \models \vartheta$ as $\psi$ holds continuously from $t_{n,d}$ onwards and no prefix of $\pi$ of length $< d$ forms a word in $\mathcal{L}(\mathcal{A})$. Otherwise, i.e. if $\mathcal{S}_{n',k}, s_{n',d} \not\models \psi$ then there must be a $d' < d$ such that $\mathcal{S}_{n',k}, s_{n',d'} \models \varphi$. Again, by induction hypothesis, $\mathcal{T}_{n,k}, s_{n,d'} \models \varphi$ and hence the same path $\pi$ as before witnesses $\mathcal{T}_{n,k}, t_{n,0} \models \vartheta$. This shows that if $\mathcal{S}_{n',k}, s_{n',0} \models \vartheta$ then $\mathcal{T}_{n,k}, t_{n,0} \models \vartheta$. The other cases, where $\mathcal{S}_{n',k}, s_{n',r} \models \vartheta$ for some $0 < r \leq k+1$, and the converse direction are proved analogously.

**Suppose** $\vartheta := \mathtt{E}(\varphi \mathtt{U}^{\mathcal{A}} \psi)$**.** Let $n, n', k \geq 0$ be such that $k > ad(\varphi)$ and $n, n' > td(\varphi)$. Again we only consider the case where $\mathcal{S}_{n',k}, s_{n',0} \models \vartheta$ and show that $\mathcal{T}_{n,k}, t_{n,0} \models \vartheta$. All other cases can be proved analogously.

We first establish the following claim which follows from a straight forward application of the induction hypothesis.

*Claim 3.* Let $\chi \in \mathrm{CTL}[\mathfrak{A},\mathfrak{B}]$ be a formula such that $td(\chi) \leq td(\psi) \leq td(\vartheta) - 1$ and $ad(\chi) \leq ad(\psi) \leq ad(\vartheta) < k$ and let $n \geq r, r' > td(\chi)$ and $n' \geq r'' > td(\chi)$. Then

$$\begin{aligned} \mathcal{T}_{n,k}, t_{r,d} \models \chi &\iff \mathcal{T}_{n,k}, t_{r',d} \models \chi \\ &\iff \mathcal{S}_{n',k}, s_{r'',d} \models \chi \end{aligned}$$

for all $0 \leq d \leq k+1$.

Let $\pi' := v_0 v_1 \ldots \models \varphi \mathtt{U}^{\mathcal{A}} \psi$ be a path witnessing that $\mathcal{S}_{n',k}, s_{n',0} \models \vartheta$. Hence, there is a $j \geq 0$ such that $\mathcal{S}_{n',k}, v_j \models \psi$ and $\mathcal{S}_{n',k}, v_i \models \varphi$ for all $0 \leq i < j$. Let $v_j := s_{r,d}$ for some $r, d$. Suppose first that $r \leq td(\vartheta) < \min\{n, n'\}$. Hence, $\mathcal{S}_{n',k}, s_{r,d'} \models \varphi$ for all $0 \leq d < d'$ and $\mathcal{S}_{n',k}, s_{r',d'} \models \varphi$ for all $0 \leq d' \leq k+1$ and $r < r' \leq n'$. By applying Claim 1, Claim 3 and the induction hypothesis we get that

– $\mathcal{T}_{n,k}, t_{r,d} \models \psi$
– $\mathcal{T}_{n,k}, t_{r,d'} \models \varphi$ , for all $d' < d$, and
– $\mathcal{T}_{n,k}, t_{r,d'} \models \varphi$ for all $n \geq r' > r$ and all $0 \leq d \leq k+1$.

Now we can choose any path $\pi_1$ from $t_{n,0}$ to $t_{r,d}$ in $\mathcal{T}_{n,k}$ such that $\mathcal{L}(\mathcal{A})$ contains a word of length $|\pi_1|$. Then $\pi \cdot (t_{r,d})^\omega$ (or $\pi \cdot t_{r,d} \cdot (t_{r,d+1})^\omega$ in case $d = 0$) witnesses that $\mathcal{T}_{n,k}, t_{n,0} \models \vartheta$.

Finally, suppose that $r > td(\vartheta)$. But then, by Claim 2, $\mathcal{T}_{n,k}, t_{n,d} \models \psi$ and $\mathcal{T}_{n,k}, t_{n,d'} \models \varphi$ for all $d' < d$. Hence, if we choose a path $\pi_1 := t_{n,0} \cdot (t_{n,1})^m \cdots t_{n,d}$ such that there is a word in $\mathcal{L}(\mathcal{A})$ of length $|\pi_1|$, then $\pi_1 \cdot (t_{n,d})^\omega$ (or $\pi_1 \cdot t_{n,d} \cdot (t_{n,d+1})^\omega$ if $d = 0$) witnesses that $\mathcal{T}_{n,k}, t_{n,0} \models \vartheta$. This concludes the proof. □

Fairness can be expressed by $\Delta\mathcal{A}_{\mathsf{fair}}$, where $\mathcal{A}_{\mathsf{fair}}$ is the standard Büchi automaton over some alphabet containing a test predicate $q?$ that recognises the language of all infinite paths on which infinitely many states satisfy $q$.

**Corollary 4.4.** *1. For all* $\mathfrak{A}, \mathfrak{B}$*:* $\mathrm{CTL}^* \not\leq \mathrm{CTL}[\mathfrak{A},\mathfrak{B}]$.
*2. There are no* $\mathfrak{A}$,$\mathfrak{B}$ *such that any* $\mathrm{CTL}[\mathfrak{A},\mathfrak{B}]$ *is equivalent to the* $\Delta\mathrm{PDL}^?[\mathrm{REG}]$ *formula* $\Delta\mathcal{A}_{\mathsf{fair}}$.

Finally, we provide some model-theoretic results which will also allow us to separate some of the logics with respect to expressive power. Not surprisingly, CTL[REG] has the finite model property which is a consquence of its embedding into $\Delta\mathrm{PDL}^?[\mathrm{REG}]$. It is also not hard to bound the size of such a model given that $\Delta\mathrm{PDL}^?[\mathrm{REG}]$ has the small model property of exponential size.

**Proposition 4.5.** *Every satisfiable* CTL[REG] *formula has a finite model. In fact, every satisfiable* CTL[NFA,DFA]*, resp.* CTL[NFA,NFA] *formula has a model of at most exponential, resp. double exponential size.*

We show now that the bound for CTL[NFA] cannot be improved.

**Theorem 4.6.** *There is a sequence of satisfiable* CTL[NFA]*-formulas* $(\psi_n)_{n\in\mathbb{N}}$ *such that the size of any model of* $\psi_n$ *is at least doubly exponential in* $|\psi_n|$.

*Proof.* Fix an even number $n > 0$. Let $[n]{:=}\{1,\ldots,n\}$. Let $\mathcal{A}$ be the following NFA over the alphabet $\Sigma{:=}\{-n,\ldots,-1,\#,1,\ldots,n\}$.

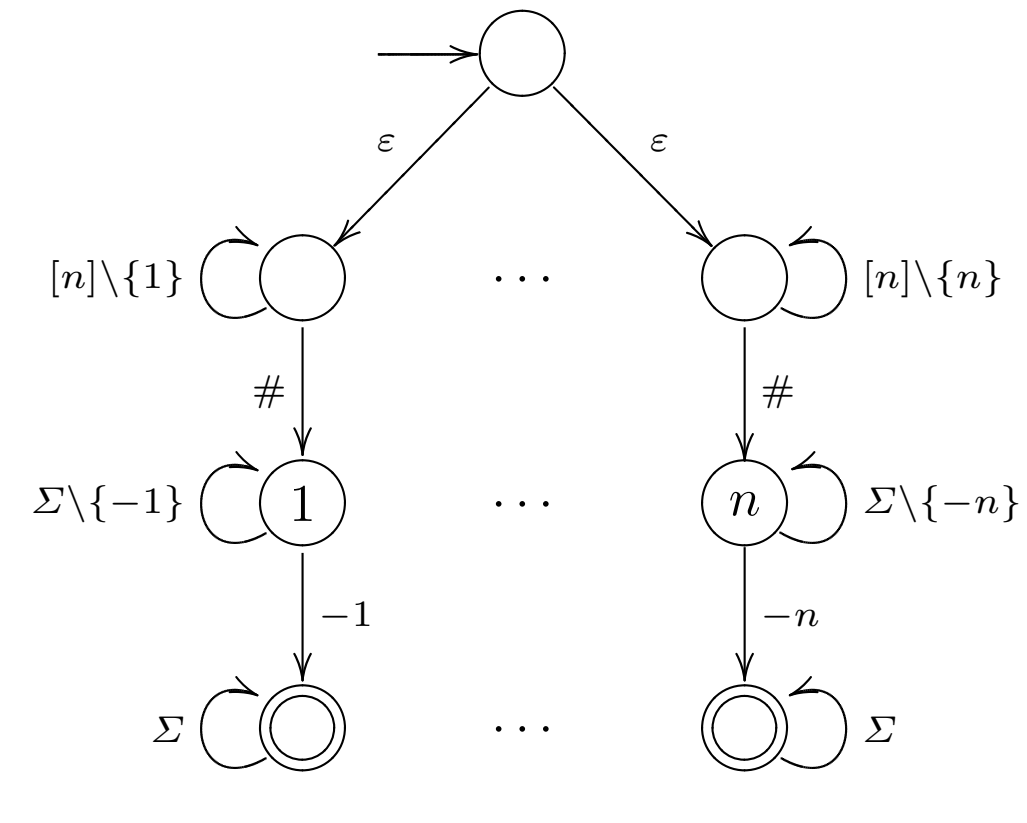

Let $Q \supset [n]$ be set of its states. The $\epsilon$-transition can be eliminated with a linear overhead. However, the $\epsilon$-transitions are more convenient for presentation purposes. In any case, the size of $\mathcal{A}$ is linear in $n$. Let $\mathcal{D}$ be a deterministic automaton for $\mathcal{A}$ obtained from the standard powerset construction [26]. Although we do not use $\mathcal{D}$ explicitly, it allows us to say that at a node of a model there is a proof obligation for $\mathtt{AF}^{\mathcal{A}}_$ in a state $S \subseteq Q$, for instance.

Let $S \subseteq [n]$. Consider a Hintikka model of some formula $\varphi$ and let $\mathtt{AF}^{\mathcal{A}}p$ occur in some node. Suppose that we have the control over the formulas $\varphi$, or over the Hintikka model, respectively. Now, we can set up $\mathcal{A}$ with the set $S$ as follows. Let $[n] \setminus S = \{s_1, \ldots, s_\ell\}$. Consider a path $\pi$ passing the labels $s_1, \ldots, s_\ell, \#$ such that along the path $p$ does not hold. At the end of this path, there is a proof obligation for $\mathtt{AF}^{\mathcal{A}}p$ in the state $S$ (w.r.t to $\mathcal{D}$). Iterating this construction with different sets $S$ yields to many proof obligations for the $\mathtt{AF}^{\mathcal{A}}$ along the iteration.

As for the lower bound, we construct a formula $\varphi$ polynomially sized in $n$ such that any of its tree model consists of two phases. The first one creates exponential many proof obligations for some instances $\mathtt{AF}^{\mathcal{A}}p$ along the path. There are doubly exponential many such paths. In the second phase the model satisfies these obligations but it also materializes all the obligations. The set of proof obligations will be so that the materialization is characteristic for this set. This property prevents any model from sharing the different materializations. To be more precise, the first phase is built from smaller blocks, called S-blocks. For each set $S \subseteq [n]$ of size $n/2$ there is a leaf such that the block imposes an additional proof obligation for $\mathtt{AF}^{\mathcal{A}}p$ in the state $S$. The first phase consists of $b := \binom{n}{n/2}/2$ many[1] layers of S-blocks. For each list $\boldsymbol{S} := S_1, \ldots, S_{n/2}$ with each element in $\binom{[n]}{n/2}$, there is a path (starting from the root) which reaches the second phase and which has collected proof obligations for $\mathtt{AF}^{\mathcal{A}}p$ in the state $S_i$ for any $i \in [n/2]$. In the last phase, the model can pick out $b = \binom{n}{n/2} - b$ sets in $\binom{[n]}{n/2}$. Only for these sets the model has a path. For a set $\{a_1, \ldots, a_{n/2}\} \in \binom{[n]}{n/2}$, the path touches the labels $-a_1, \ldots, -a_{n/2}$ in some order. The node after the last label has no successor, and it is the only state on the path at which $p$ holds. The passed labels transform the proof obligations. The only node which can fulfill the the proof obligations is a dead-end node. The combination of both properties implement the said materialization. This final phase is implemented by a so-called T-block.

The encoding of this paradigm uses two kinds of counters: one to iterate the S-blocks, and the others to control the branching in any T-block. We write $\boldsymbol{C}$ for a list of $n$ (distinct) propositions which are intended to be used as $n$-bit counter.

Let $\boldsymbol{A}$, $\boldsymbol{B}$, $\boldsymbol{C}$—possibly indexed—be counters, $\ell \in \mathbb{N}$, $v \in \{0, \ldots, 2^n - 1\}$, and $\Delta \subseteq \Sigma$. There are CTL-formulas of polynomial size (in $n$) which encode the following properties.

[1] Indeed, $\binom{n}{n/2} = \frac{(n-1)! \cdot n}{(n/2)!(n/2-1)! \cdot n/2} = 2\binom{n-1}{n/2}$ is even.

| Formula | Property |
|---|---|
| $\ulcorner \boldsymbol{C} = v \urcorner$, $\ulcorner \boldsymbol{C} \neq v \urcorner$ | The counter $\boldsymbol{C}$ has (not) the value $v$. |
| $\ulcorner \mathtt{AX}^{\Delta} \boldsymbol{A} = \boldsymbol{B} \urcorner$ | The value of $\boldsymbol{A}$ in any $\Delta$-successor is the value of $\boldsymbol{B}$ of the current state. |
| $\ulcorner \mathtt{AX}^{\Delta} \boldsymbol{A} = \boldsymbol{B} + 1 \urcorner$ | The value of $\boldsymbol{A}$ in any $\Delta$-successor is the successor value of $\boldsymbol{B}$ of the current state. If $\boldsymbol{B}$ represents $2^n$ the behavior is undefined. |
| $\ulcorner \boldsymbol{A} = \sum_{i=0}^{\ell} \boldsymbol{B}_i \urcorner$ | The value of $\boldsymbol{A}$ is the sum of the values of $\boldsymbol{B}_i$ for all $i = 0 \ldots \ell$. Here, we allow (polynomial many) additional counters, respectively variables, to compute the sum successively. |

The final formula $\varphi$ uses the propositions $p$, and the counters $\boldsymbol{C}$ and $\boldsymbol{C}_i$ for $i = 0, \ldots, n$.

*Encoding of S-blocks.* For $\Delta \subseteq \Sigma$ and $\psi$ a CTL-formula, the formula

$$!\mathtt{X}^{\Delta}\psi := \mathtt{AX}^{\Sigma \setminus \Delta}\mathtt{ff} \ \wedge \ \mathtt{EX}^{\Delta}\mathtt{tt} \ \wedge \ \mathtt{AX}^{\Delta}\psi$$

forces that for any of its models there are only $\Delta$-successors and at each of them $\psi$ holds. Note that for an $a \in \Delta$ there might be more than one $a$-successors.

The enumeration of all $S \in \binom{n}{n/2}$ is constructed level by level. An element $S$ is enumerated increasingly. Thereto, the auxiliary formulas $\varphi_{m,\ell}$ are introduced for $\ell$ the number of levels remaining and $m$ the maximal number seen along an enumeration so far.

$$\begin{aligned} \varphi_{m,0} &:= !\mathtt{X}^{\{\#\}}\mathtt{tt} \\ \varphi_{m,\ell} &:= !\mathtt{X}^{\{m+1,\ldots,n+1-\ell\}}\neg p \qquad \text{if } \ell > 0 \end{aligned}$$

Finally, an S-block is forced by

$$\sigma := \mathtt{AF}^{\mathcal{A}}p \ \wedge \ \neg p \ \wedge \ \varphi_{0,n/2} \ \wedge \ \bigwedge_{\substack{m \in [n] \\ k \in [n/2]}} \mathtt{AX}^{\Sigma^{k-1}\{m\}}\varphi_{m,n/2-k}.$$

Any (tree) model of $\sigma$ enumerates all subsets of $[n]$ of size $n/2$, and ensures that along the enumeration $p$ does not hold while the proof obligation $\mathtt{AF}^{\mathcal{A}}p$ is imposed on the root. That is, for any sequence $a_1, \ldots, a_{n/2+1}$ in $\Sigma$ the following properties are equivalent.

- $a_1, \ldots, a_{n/2}$ is a strictly increasing sequence in $[n]$, and $a_{n/2+1} = \#$.
- there exists a path $s_0, a_1, s_1, a_2, s_2, \ldots$ starting at $s$ such that $s_i \models \neg p$ for all $i \in \{0, \ldots, n/2\}$, and $s_0 \models \mathtt{AF}^{\mathcal{A}}p$.

*Encoding of T-blocks.* An T-block is a tree with $b$ leaves. The encoding is similar to that of an S-block. Additionally, at each node $v$ we use a counter $C_0$ and counters $C_i$ for each outgoing label $-i$. The counter $C_0$ contains the number of leaves of the tree[2] at $v$. Similarly, $C_i$ stands for the number of leaves at the respective subtree. The counters $C_i$ must sum up to $C_0$. In analogy to $\varphi_{m,\ell}$, each formula $\psi_{m,\ell}$ is responsible for a certain level. However, the expression $!\mathtt{X}^{\Delta}$ is replaced by a variation additionally depending on the counter $\boldsymbol{C}_i$.

$$
\begin{aligned}
\psi_{m,0} &:= \ulcorner \boldsymbol{C}_0 = 1 \urcorner \wedge p \wedge \mathtt{AX}\mathtt{ff} \\
\psi_{m,\ell} &:= \neg p \wedge \bigwedge_{a \in \Sigma \setminus \{-n,\dots,-1\}} \mathtt{AX}^{\{a\}}\mathtt{ff} \\
&\quad \wedge \bigwedge_{i=m+1}^{n+1-\ell} \Big\{ \Big( \ulcorner \boldsymbol{C}_i \neq 0 \urcorner \leftrightarrow \mathtt{EX}^{\{-i\}}\mathtt{tt} \Big) \\
&\qquad\qquad \wedge \ulcorner \mathtt{AX}^{\{-i\}} \boldsymbol{C}_0 = \boldsymbol{C}_i \urcorner \Big\}
\end{aligned}
$$

A T-block is represented by the formula $\tau$ defined as

$$
\ulcorner \boldsymbol{C}_0 = b \urcorner \wedge \psi_{0,n/2} \wedge \bigwedge_{\substack{m \in [n] \\ k \in [n/2]}} \mathtt{AX}^{\Sigma^{k-1}\{-m\}} \psi_{m,n/2-k}
$$

*Encoding.* Now, the S-blocks can be iterated $b$-times.

$$
\begin{aligned}
\varphi := \ulcorner \boldsymbol{C} = 0 \urcorner \wedge \mathtt{AG}^{\Sigma^*} \Big( & \ulcorner \mathtt{AX}^{\Sigma \setminus \{\#\}} \boldsymbol{C} = \boldsymbol{C} \urcorner \wedge \\
& \ulcorner \mathtt{AX}^{\{\#\}} \boldsymbol{C} = \boldsymbol{C} + 1 \urcorner \Big) \\
\wedge\ & \mathtt{AG}^{\Sigma^*\{\#\}} \left( \ulcorner \boldsymbol{C} \neq b \urcorner \to \sigma \right) \\
\wedge\ & \mathtt{AG}^{\Sigma^*\{\#\}} \left( \ulcorner \boldsymbol{C} = b \urcorner \to \tau \right)
\end{aligned}
$$

*$\varphi$ is satisfiable.* We construct a tree model of $\varphi$. Obviously, the existence of the first phase—as mentioned in the introductive text—is guaranteed because $\tau$ and $\varphi$ without its last conjunct have bisimilar models only. Given a path $\pi$ from the root to the last element of the first phase, it remains to show how to continue with a T-blocks. By the construction of $\sigma$ and $\varphi$, there are sets $S_1, \dots, S_b \in \binom{[n]}{n/2}$ such that the path has collected only proof obligation of $\mathtt{AF}^{\mathcal{A}}p$ for the states $S_1$ to $S_b$. Let $\mathcal{S} := \{S_i \mid i \in [b]\}$. Now, set

$$
\mathcal{T} := \left\{ T \in \binom{[n]}{n/2} \mid [n] \setminus T \notin \mathcal{S} \right\}.
$$

Note that $|\mathcal{T}| = \binom{n}{n/2} - |\mathcal{S}| \geq \binom{n}{n/2} - b = b$. Choose a subset $\mathcal{T}' \subseteq \mathcal{T}$ of size $\binom{n}{n/2} - b$. The formula $\tau$ forces $b$ branches. Therefore, for each $T \in \mathcal{T}'$ we construct a branch which passes the labels $-t_1, \dots, -t_b$, where $t_1, \dots, t_b$ is an increasing enumeration of $T$. For any $S \in \mathcal{S}$, the sets $S$ and $T$ are not disjoint.

[2] Because CTL is bisimilar, there might be more than one out-going edge with a given label $a \in \Sigma$. In this case, we pick out one such edge. So, the term “tree” refers to the tree thinned out.

Indeed, if they are disjoint then $[n] \setminus T = S$ as both have the same size $n/2$. But this is contradiction to $T \in \mathcal{T}$. The non-disjointness ensures that any proof obligation in $S$ is turned into an obligation for a set of states containing a final state, after passing the labels $-t_1, \ldots, -t_b$. However, this state models $p$, and hence all proof obligations disappear.

*Lower bound.* Consider a model $\mathcal{T}$ of $\varphi$. Because $\binom{2k}{k} \geq 2^k$ for any $k \in \mathbb{N}$, the set $\binom{\binom{[n]}{n/2}}{b}$ has at least doubly exponential size in $n$. For any set $\mathcal{S} \in \binom{\binom{[n]}{n/2}}{b}$ there is a rooted path $\pi_{\mathcal{S}}$ through the S-blocks of $\mathcal{T}$ which got proof obligations for $\mathtt{AF}^{\mathcal{A}}p$ for every $S \in \mathcal{S}$ and ends at the first node of a T-block. Let $\mathcal{S}$ and $\mathcal{S}'$ two different sets in $\binom{\binom{[n]}{n/2}}{b}$. As for the lower bound, it suffices to show that the last nodes of $\pi_{\mathcal{S}}$ and $\pi_{\mathcal{S}'}$ are different. Assume that they are identical. The T-block starting at the last node shows $b$ branches, each naming (the negative of each element of) a set $T \subseteq [n]$ of size $n/2$. As in the case of satisfiability, the proof obligations got transformed by each branch. Since a T-block is a dead end, a transformed proof obligation must refer to a set which contains a final state of $\mathcal{A}$. Therefore, $T$ must intersect with any element of $\mathcal{S} \cup \mathcal{S}'$. That is, $([n] \setminus T) \notin \mathcal{S} \cup \mathcal{S}'$. In total, each of the $b = \binom{n}{n/2} - b$ branches names a different set which is not in $\mathcal{S} \cup \mathcal{S}'$. So, $|\mathcal{S} \cup \mathcal{S}'| = b$. Being of size $b$, both $\mathcal{S}$ and $\mathcal{S}'$ are identical. Contradiction. □

The next theorem provides information about the type of models we can expect – which is useful for synthesis purposes.

**Theorem 4.7.** *1. There is a satisfiable* CTL[VPL] *formula which does not have a finite model.*

*2. There is a satisfiable $\varphi \in$* CTL[DCFL] *s.t. no pushdown system is a model of $\varphi$.*

*3. Every satisfiable* CTL[VPL] *formula has a model which is a visibly pushdown system.*

We prove the first two parts here. Due to its length, the proof of Part 3 is given as a separate lemma (Lemma 4.8) below.

*Proof (Part 1. and 2.).*

1. Consider a visibly pushdown alphabet with some push-action $a$ and a pop-action $b$. Let $L := \{a^n b^n \mid n \geq 1\}$ which is known to be a VPL. Note that VPL are closed under complement and under intersections with regular languages. Let

$$\varphi := \mathtt{EF}^a \mathtt{AG}^{a^*} \mathtt{EF}^a \mathtt{tt} \wedge \mathtt{AG}^L \mathtt{AX} \mathtt{ff} \wedge \mathtt{AG}^{\overline{L} \cap a^* b^*} \mathtt{EF}^b \mathtt{tt}$$

The first conjunct requires an infinite $a$-path to exist. The second states that every path of the form $a^n b^n$ ends in a state without successor, and the third one says that every path of the form $a^n b^m$ with $m < n$ can be extended by at least another $b$-action. It is not hard to see that $\varphi$ cannot be satisfied in a finite model. Furthermore, $\varphi$ is satisfiable, for instance in the following model.

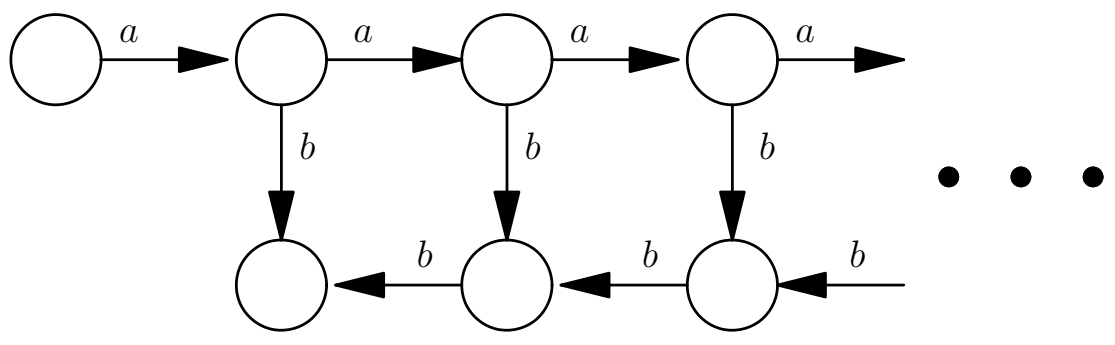

2. Note that the following languages are DCFLs: $L_1 := \{a^n b^m \mid 1 \leq m < n\}$, $L_2 := \{a^n b^n \mid n \in \mathbb{N}\}$, $L_3 := \{a^* b^n c^m \mid 1 \leq m < n\}$, and $L_4 := \{a^* b^n c^n \mid n \in \mathbb{N}\}$. Now we construct a formula $\varphi \in$ CTL[DCFL] whose models are bisimilar to the following transition system.

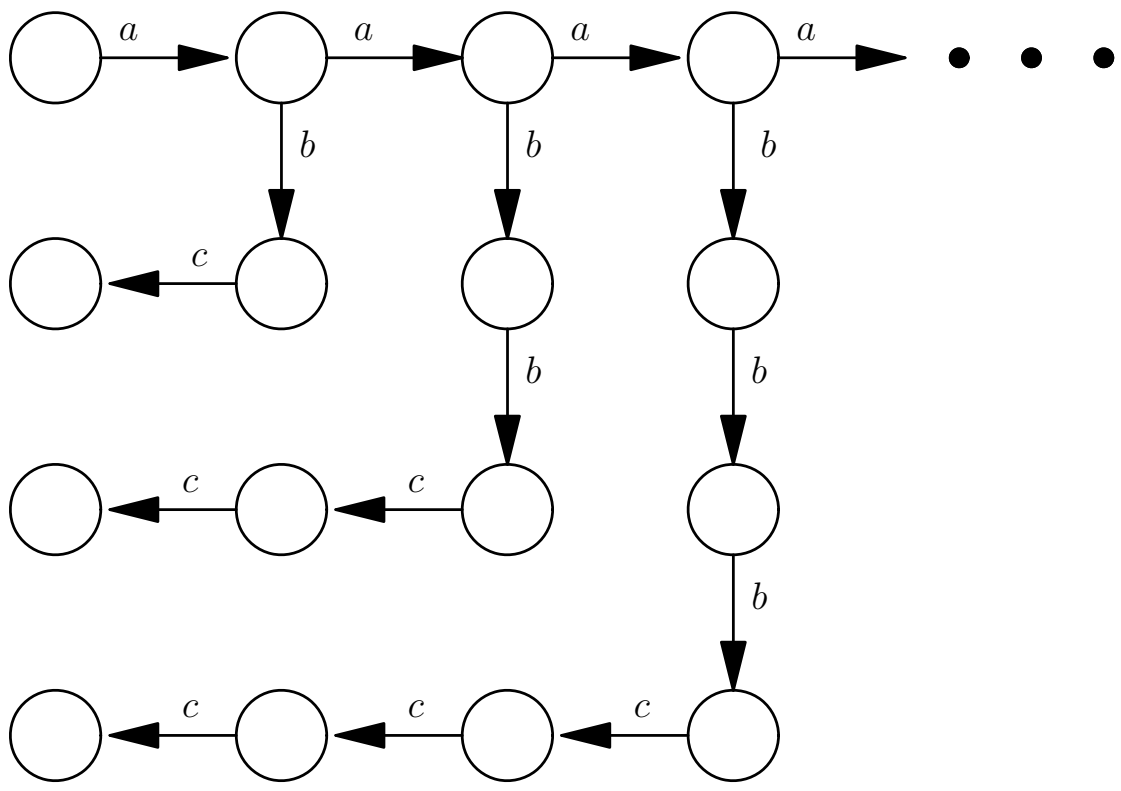

It is defined as follows.

$$\begin{aligned}\varphi \;:=\;& \mathtt{AG}^{b\cup c}\mathtt{ff} \wedge \mathtt{EF}^{a}\mathtt{AG}^{a^*}(\mathtt{EF}^{a}\mathtt{tt} \wedge \mathtt{EF}^{b}\mathtt{tt})\\ \wedge\;& \mathtt{AG}^{L_1}(\mathtt{EF}^{b}\mathtt{tt} \wedge \mathtt{AG}^{a\cup c}\mathtt{ff}) \wedge \mathtt{AG}^{L_2}\mathtt{EF}^{c}\mathtt{tt}\\ \wedge\;& \mathtt{AG}^{L_3}(\mathtt{EF}^{c}\mathtt{tt} \wedge \mathtt{AG}^{a\cup b}\mathtt{ff}) \wedge \mathtt{AG}^{L_4}\mathtt{AX}\mathtt{ff}\end{aligned}$$

The first line requires an infinite $a$-path with $b$-successors for every state apart from the first one. The second line requires every path of the form $a^n b$ to be followed by paths of the form $b^{n-1}c$ and nothing else, the third line requires every path of the form $a^* b^n c$ – which then has to be in fact $a^n b^n c$ – to be followed by $c^{n-1}$ ending in a deadlock state.
Note that such a model cannot be represent as a pushdown model because the set of all maximal paths in a pushdown model forms a context-free language but $\{a^n b^n c^n \mid n \in \mathbb{N}\}$ is known not to be a CFL. □

**Lemma 4.8.** *Every satisfiable* CTL[VPL] *formula has a model which is a visibly pushdown system.*

*Proof.* Beforehand, we harmonize the definitions of two kinds of automata, and of a push down system.

Let $\Sigma = (\Sigma_{\mathtt{c}}, \Sigma_{\mathtt{r}}, \Sigma_{\mathtt{i}})$ be a pushdown alphabet [2]. For the following three definitions, $Q$ refers to a set of states, $q_0 \in Q$ to an initial state, $\Gamma$ to a stack alphabet containing the bottom-of-stack symbol $\bot$, and $col : Q \to \mathbb{N}$ to a function coloring the states $Q$. Moreover, we implicitly use the standard [2, 19] notations of a configuration, and of a run on $\omega$-words over $\Sigma$ and on infinite trees over $\Sigma$, respectively. For simplicity, let $T$ be the set $(Q \times \Sigma_{\mathtt{c}} \times (\Gamma \setminus \{\bot\} \times Q)^*) \cup (Q \times \Sigma_{\mathtt{i}} \times Q^*) \cup (Q \times \Sigma_{\mathtt{r}} \times \Gamma \times Q^*)$. We write $\langle (q_1, B_1), \ldots (q_n, B_n) \rangle$ for an element in $(\Gamma \setminus \{\bot\} \times Q)^*$. A *ordered visibly pushdown system* (oVPS) over $\Sigma$ is a tuple $P = (Q, \Gamma, \delta, q_0)$ such that $\delta \subseteq T$ and $\delta$ is deterministic. An oVPS $P$ induces an $\Sigma$-labeled and ordered tree by unrolling $\delta$. A *parity tree automaton* over $\Sigma$ is a tuple $\mathcal{A} = (Q, \delta, q_0, col)$ such that $\delta \subseteq Q \times \Sigma \times Q^*$. A *stair parity visibly pushdown tree automaton* [24] over $\Sigma$ is a tuple $\mathcal{A} = (Q, \Gamma, \delta, q_0, col)$ such that $\delta \subseteq T$. Any such automaton is said to be *satisfiable* if there exists a tree which it accepts.

Given a stair parity VPTA $\mathcal{A}$, we construct an oVPS such that its induced tree is accepted by $\mathcal{A}$. As for the claim of Thm. 4.7 Part 3, for any $\Delta$PDL$^?$[VPA]- and any CTL[VPA]-formula $\varphi$ there is a stair parity VPTA which accepts exactly the unique diamond path and unique $\Delta$-path Hintikka tree models of $\varphi$ [23, Lem. 24]. By the announced implication there exists a oVPS which admits such a Hintikka model for $\varphi$. From this, one obtains a VPS [24] satisfying $\varphi$, as just as one gets a tree model from a Hintikka model [23, Prop. 23].

Let $\mathcal{A} = (Q^{\mathcal{A}}, \Gamma, \delta^{\mathcal{A}}, q_0^{\mathcal{A}}, col^{\mathcal{A}})$ be a stair parity visibly pushdown tree automaton over a a pushdown alphabet $\Sigma = (\Sigma_{\mathtt{c}}, \Sigma_{\mathtt{r}}, \Sigma_{\mathtt{i}})$.

**Definition 4.9.** *Wlog.* $col^{\mathcal{A}} \colon Q^{\mathcal{A}} \to \mathbb{N} \setminus \{0\}$, *and* $Q^{\mathcal{A}} = \{1, \ldots, |Q^{\mathcal{A}}|\}$. *The parity tree automaton* $\mathcal{B} := (Q^{\mathcal{B}}, \delta^{\mathcal{B}}, q_0^{\mathcal{B}}, col^{\mathcal{B}})$ *is defined as follows.*

- $Q^{\mathcal{B}} := \left(Q^{\mathcal{A}} \times \Gamma \times 2^{Q^{\mathcal{A}}}\right) \dot\cup \{\checkmark\}$.
- $q_0^{\mathcal{B}} := (q_0^{\mathcal{A}}, \bot, \emptyset)$.
- $col^{\mathcal{B}}\big((q, _, _)\big) := col^{\mathcal{A}}(q)$ *for all* $q \in Q^{\mathcal{A}}$, *and* $col^{\mathcal{B}}(\checkmark) := 0$.

*The relation* $\delta^{\mathcal{B}}$ *is given by case distinction on* $\Sigma$.

**Always:** $(\checkmark, a, \langle \checkmark \rangle) \in \delta^{\mathcal{B}}$ *for all* $a \in \Sigma$.

**For all** $a \in \Sigma_{\mathtt{i}}$ **and** $(q, a, \langle q_1, \ldots, q_k \rangle) \in \delta^{\mathcal{A}}$**:** *Then* $((q, \gamma, R), a, \langle (q_1, \gamma, R), \ldots, (q_k, \gamma, R) \rangle) \in \delta^{\mathcal{B}}$ *for all* $\gamma \in \Gamma$.

**For all** $a \in \Sigma_{\mathtt{r}}$ **and** $(q, \gamma, a, \langle q_1, \ldots, q_k \rangle) \in \delta^{\mathcal{A}}$**:** *Then* $((q, \bot, R), a, \langle (q_1, \bot, R), \ldots, (q_k, \bot, R) \rangle) \in \delta^{\mathcal{B}}$. *And* $((q, \gamma, R), a, \langle \checkmark \rangle) \in \delta^{\mathcal{B}}$ *if* $q_i \in R$ *for all* $i = 1, \ldots, k$.

**For all** $a \in \Sigma_{\mathtt{c}}$ **and** $(q, a, \langle (\gamma_1, q_1), \ldots, (\gamma_k, q_k) \rangle) \in \delta^{\mathcal{A}}$**:** *Let* $R_1, \ldots, R_k \subseteq Q^{\mathcal{A}}$ *be arbitrary. Then* $((q, \gamma', R'), a, \langle \boldsymbol{w}_1 \ldots \boldsymbol{w}_k \rangle) \in \delta^{\mathcal{B}}$ *where* $\boldsymbol{w}_i$ *for* $i = 1, \ldots, k$ *is a vector over* $Q^{\mathcal{B}}$ *of length* $1+|Q^{\mathcal{A}}|$. *Its first component is* $(q_i, \gamma_i, R_i)$, *followed by* $(r, \gamma', R')$ *if* $r \in R_i$, *or by* $\checkmark$ *otherwise, for all* $r \in Q^{\mathcal{A}}$ *increasingly.*

Note that from any transition in $\mathcal{B}$ its generating transition in $\mathcal{A}$ can be reconstructed.

**Lemma 4.10.** *If* $\mathcal{A}$ *is satisfiable then so* $\mathcal{B}$ *is.*

*Proof.* Suppose that $\mathcal{A}$ accepts a tree $t_{\mathcal{A}}$. Let $t'_{\mathcal{A}}$ be the tree $t_{\mathcal{A}}$ but additionally annotated with configurations of $\mathcal{A}$ witnessing that $t_{\mathcal{A}}$ is accepted by $\mathcal{A}$. Starting from the root, the tree $t'_{\mathcal{A}}$ is successively rearranged to a tree $t_{\mathcal{B}}$ accepted by $\mathcal{B}$. Let a node $v$ be given. If at $v$ the automaton $\mathcal{A}$ does an internal operation or a pop operation then this nodes remains. Now, assume that $\mathcal{A}$ does a push operation along $v$ to a child $w$. Consider the occurrences of all pop operations corresponding to the push operation from $v$ to $w$ on all branches arising from $w$. Let $R$ be the states reached by $\mathcal{A}$ as a result of the exhibited pop operation. Hence, for any $r \in R$ there is a subtree $t_r$ below $w$ annotated with the state $r$. For all $r \in Q^{\mathcal{A}}$, increasingly, the node $v$ got the following subtree as a sibling. If $r \in R$ then we take $t_r$ and otherwise some (infinite) tree. The new sibling are inserted right after $v$ and a head of its siblings in the first place.

The construction ensures that the resulting tree is accepted by $\mathcal{B}$. Indeed, let $\pi$ be a path starting in $q_0^{\mathcal{B}}$. If $\pi$ touches $\checkmark$, it keeps doing so. Hence, the path is accepted. Otherwise, the path corresponds to a branch in $\mathcal{A}$ where the immediate run corresponding to a maximally matching word [2] are omitted. For each such word, a branch is forked, cf. the first component of the $w_i$s in Def. 4.9. Hence, $\pi$ corresponds to a branch in $\mathcal{A}$. However, the positions of the maximally matching words are not taken into account for the acceptance condition. But, this restriction is just the stair parity condition. Hence, $\pi$ is accepted.

**Definition 4.11.** *Let $\mathcal{C}$ be a parity tree automaton over $\Sigma$ with states $Q$ and transitions $\delta$. A triple $(V, E, r, \ell)$ is a* finite interpretation *for $\mathcal{C}$ iff $V$ is a finite set of nodes, $E : V \to V^+$ is a successor function with ordered children, $r \in V$ is its root, and $\ell\colon V \to (Q \times \Sigma)$ is a labeling function which in conform with $\mathcal{C}$. That is, $E(v_0) = (v_1, \ldots, v_n)$ and $\ell(v_i) = (q_i, a_i)$ for all $i \in \{0, \ldots, n\}$ imply $(q_0, a_0, (a_1, \ldots, q_n)) \in \delta$, for any $v_0, \ldots, v_n \in V$, $q_0, \ldots, q_n \in Q$, and $a_0, \ldots, a_n \in \Sigma$. Such a finite interpretation is a* finite model *of $\mathcal{C}$ iff $\mathcal{C}$ accepts the tree resulting from unrolling $(V, E, r, \ell)$ at its root. The labels of this tree follow the $\Sigma$-part of $\ell$.*

**Theorem 4.12.** *Any satisfiable parity tree automaton has a finite model.*

*Proof.* The emptiness problem can be reduced to the question whether or not the automaton player has a winning strategy for a finite parity game [19]. The set of winning position is computable. Hence, fixing one outgoing edge of a position of the automaton player leads directly to the claimed graph.

Finally, the translation in Def. 4.9 and the reduction in Lem. 4.10 can be reversed.

**Definition 4.13.** *Let $G = (V, E, r, \ell)$ a finite model of $\mathcal{B}$. Then $G$ induces an oVPS $P := (V, \Gamma^P, \delta^P, r)$, where the stack alphabet $\Gamma^P$ is $(Q \to V) \dot\cup \{\bot\}$. The transition relation $\delta^P$ is given as follows. Let $v \in V$ be labeled with $(_, a) \in Q \times \Sigma$. For any $a \in \Sigma_{\mathtt{i}}$, $\delta^P$ contains $(v, a, E(v))$. And for any $a \in \Sigma_{\mathtt{r}}$, $\delta^P$ contains $(v, a, \bot, E(v))$ and $(v, a, \rho, \rho(v))$ for any function $\rho : Q \to V$. As for the push operations, let $E(v) = \boldsymbol{v}_1 \ldots \boldsymbol{v}_k$ and let $\boldsymbol{v}_i = v_{i,0}, \ldots, v_{i,|Q|}$ for each $i$, due to the conformity of $G$ with $\mathcal{B}$. Then $\delta^p$ contains $(v, a, ((\rho_1, v_{1,0}), \ldots, (\rho_k, v_{k,0})))$ where $\rho_i : Q \to V$ is some (fixed) function such that $\rho_i(q) = v_{i,q}$ if the $\Sigma$-part of $\ell(v_{i,q})$ is not $\checkmark$.*

Because, in the tree resulting from unrolling $G$, no rooted branch reaches the state $\checkmark$, transitions leaving this state need not be translated.

**Theorem 4.14.** *Let $G$ be a finite model of $\mathcal{B}$. Then the oVPS $P$ is a model of $\mathcal{A}$.*

*Proof.* In the unrolled tree of $P$, any maximal path $\pi$ which starts at the root is infinite, following the labeling function. Analogously to the proof of Lem. 4.10, such a path meets the stair parity condition. Indeed, it suffices to consider the interrupted path which skips the minimally matching words in the factorization of (the word labeling) $\pi$. Such an interrupted path corresponds to a path in $G$ meeting the parity condition of $\mathcal{B}$. Hence, $\pi$ fulfills the stair parity condition for $\mathcal{A}$.

As for the underdetermination of the functions $\rho_i$ in the case $\Sigma_{\mathtt{c}}$: if the $\Sigma$-part of $\ell(v_{i,q})$ is $\checkmark$, the value of $\rho_i(q)$ is irrelevant as the function will be never evaluated at $q$—as long as only rooted paths are considered. This is ensured by the condition "$q_i \in R$" in the case $\Sigma_{\mathtt{r}}$ of Def. 4.9 and by the conformity of $G$ with $\mathcal{B}$.

This completes the proof of Lemma 4.8 and therefore Thm. 4.7 Part 3. □

Putting Prop. 4.5 and Thm. 4.7 together we obtain the following separations. We expect that CTL[DCFL] $\lneq$ CTL[CFL] also holds but have no formal proof at the moment. Note that the corollary can also be obtained from language theoretical observations.

**Corollary 4.15.** CTL[REG] $\lneq$ CTL[VPL] $\lneq$ CTL[DCFL].

*Proof (Alternative proof).* Here we give an alternative proof of the corollary, based on language theoretical observations.

We show that CTL[REG] $\lneq$ CTL[VPL] $\lneq$ CTL[DCFL]. In fact, the logics can be separated already on finite words (with the obvious modification of their semantics) and this extends in the usual way to a full separation of the logics with the standard semantics. For, on finite words, every formula of CTL[REG] translates into an equivalent formula of Monadic Second-Order Logic (MSO) and thus into a DFA. As there is a VPL which is not regular, it is trivial to construct a CTL[VPL]-formula which on finite words is not equivalent to any DFA and hence not equivalent to any CTL[REG]-formula. Similarly, a correspondence between VPL and $\mathrm{MSO}_\mu$ (see [2]) can be used to separate CTL[VPL] from CTL[DCFL]. □

## 5 Satisfiability

In this section we study the complexity of the satisfiability problem for a variety of CTL[$\mathfrak{A}$,$\mathfrak{B}$] logics. The presented lower and upper bounds, as shown in Fig. 2, also yield sharp bounds for EF[_] and CTL[_].

| complexity | DFA | NFA | DVPA | VPA | DPDA, PDA |
|---|---|---|---|---|---|
| DFA, NFA | 1 | 2 | 2 | 3 | |
| DVPA, VPA | 2 | 3 | 2 | 3 | undecidable |
| DPDA, PDA | | | undecidable | | |

**Fig. 2.** The complexities of checking satisfiability for a CTL[$\mathfrak{A}$,$\mathfrak{B}$] formula. The rows contain different values for $\mathfrak{A}$ as the results are independent of whether or not the automata from this class are deterministic. A natural number $k$ means that the respective logic is complete for $k$-EXPTIME.

**Theorem 5.1.** *The satisfiability problems for* CTL[DPDA, _] *and for* CTL[_, DPDA] *are undecidable.*

*Proof.* Harel et al. [17] show that PDL over regular programs with the one additional language $L{:=}\{a^nba^n \mid n \in \mathbb{N}\}$ is undecidable. Since $L \in \text{DCFL} \supseteq \text{REG}$, EF[DPDA] is undecidable and hence so is CTL[DPDA, _]. As for the second claim, the undecidable intersection problem of two DPDA, say $\mathcal{A}$ and $\mathcal{B}$, can be reduced to the satisfiability problem of the CTL[_, DPDA]-formula $\mathtt{AU}^{\mathcal{A}}\mathtt{AXff} \wedge \mathtt{AU}^{\mathcal{B}}\mathtt{AXff}$. □

**Theorem 5.2.** *The upper bounds for the satisfiability problem are as in Fig. 2.*

*Proof.* By Thm. 4.2(3), CTL[$\mathfrak{A}$, $\mathfrak{B}$] can be translated into $\Delta\text{PDL}^?[\mathfrak{A} \cup \mathfrak{B}]$ with a blow-up that is determined by the worst-case complexity of transforming an arbitrary $\mathfrak{A}$-automaton into a deterministic one. The claim follows using that REG $\subseteq$ VPL and that the satisfiability problem for $\Delta\text{PDL}^?$[REG] is in EXPTIME [15] and for $\Delta\text{PDL}^?$[VPL] is in 2EXPTIME [23]. □

**Theorem 5.3.** *1.* CTL[DFA, NFA] *and* CTL[_, DVPA] *are 2EXPTIME-hard.*
*2.* CTL[DVPA, NFA] *and* CTL[_, DVPA $\cup$ NFA] *are hard for 3EXPTIME.*

The reduction uses the alternating tiling problem.

**Definition 5.4.** *The* alternating tiling problem *is the following. Given a set $T$ of tiles, $H, V \subseteq T^2$, $s \in T$, $f\colon \mathbb{N} \to \mathbb{N}$, and $\alpha\colon T \to \{0,1,2\}$ such that $H \subseteq \{(t,t') \mid \alpha(t) = \alpha(t')\}$ decide whether there is a* tiling tree*. That is, a finite tree such that*

- *any node is labeled with $t_1, \ldots, t_m$ for $m{:=}f(|T|)$,*
- *$t_1 = s$ for the root,*
- *$t_i H t_{i+1}$ for all $1 \leq i < m$,*
- *the node has $\alpha(t_m)$ successors, and*
- *for each successor labeled with $t'_1, \ldots, t'_m$ holds $t_i V t'_i$ for all $1 \leq i \leq m$.*

The function $\alpha$ realizes alternation. Note that, if the range of $\alpha$ is $\{0,1\}$ the definition corresponds the usual one version for one player [30]. Therefore, we refer to a node in a tiling tree as a *row* and to its components as *columns*. So, $H$ represent the horizontal and $V$ the vertical matching relation.

To describe the complexity of alternating tiling we assume a reasonable encoding of $T$, $H$ et cetera. In particular, the function $f$ is given as a term. As we want to characterize complexity classes far beyond EXPTIME the usual corridor tiling [30] does not suffice because an explicit naming of the width would require to much space.

Combining the technique of tiling and alternation [8], we obtain the following characterization.

**Lemma 5.5.** *The class of alternating tiling problems where their functions $f$ is exponential is 2EXPTIME-complete. Similar, the restriction to doubly exponential functions is complete for 3EXPTIME.*

In Def. 5.4, the restriction on $H$ with respect to $\alpha$ is not necessary for the completeness for the respective completity class. However, it simplifies that subsequent hardness proof for CTL[DFA,NFA].

*Proof (of Thm. 5.3(1)).* Given an alternating tiling problem consisting of $T$, $H$, $V$, $s$, $f$ and $\alpha$ as in Def. 5.4 such that $f$ is exponential. Set $n{:=}|T|$, $m{:=}f(n)$ and let $m'$ be the number of bits to count from 0 to $m{-}1$, that is $n'{:=}\lfloor \log_2(m{-}1)\rfloor{+}1$. Note that $n'$ is polynomially bounded in $n$. W.l.o.g. $T = \{1, \dots, n\}$.

It is pretty easy to find a CTL-formula $\varphi$ such that any of its models looks like an tiling tree (up to bisimulation). Thereto, the tiles are encoded by propositions, say $t_1, \dots, t_n$. Any sequent of tiles in a node of the tree is represented by a chain of nodes in the model of the respective length. The length is ensured by a binary counter with $n'$ bits. In (pure) CTL all properties can specified except for the constraint on $V$. Therefore, the formula would need to look about $m$ steps into the future while have a size polynomial in $n$.

The $V$-constraint refers only to any those two immediately consecutive positions on which the counter has the same value. To bridge between those two positions, a proof obligation is created by an $\mathtt{AU}^{\mathcal{A}}$-subformula. The key idea is that for the correctness we can replace $\mathcal{A}$ by the deterministic automaton obtained from the standard powerset-construction [26]. In other words, we are allowed to construct an exponentially sized automaton but which has a small description. The mentioned obligation reflects the value of the counter and the expected tile at the second position. However, its creating requires that the outgoing edge is replaced by a chain of edges. Each edge copies another bit from the counter to the proof obligation. As long as the nodes of the model represent the same row, the programmed proof obligation are not armed, that is, they can not reach any final state. The change to the next row arms the obligations. Along the path to the second position, at every tile position an appendix in the model checks every proof obligation. If the current value of the counter does not match the stored value in the obligation the model ensures that the obligation is satisfied trivially. Otherwise, the (only remaining) obligation matches the chosen tile with the expected tile. Finally at every second change of the row, the model disposes of the proof obligations.

Formally, we will construct a formula $\varphi$ over the alphabet

$$\Sigma{:=}\{\mathtt{nextCol}, \mathtt{nextRow}, \mathtt{ifNeq}, \mathtt{then}, \mathtt{else}\} \cup \Gamma$$

where $\Gamma{:=}\{\mathtt{bit}_i^b \mid i \in [n], b \in \mathbb{B}\}$. As boolean values we use 0 and 1. The label `nextCol` separates two columns in the same row, and `nextRow` indicates a new node in the tiling tree. The set $\Gamma$ is used to program the proof obligations, which are verified with help of `ifNeq`, `then` and `else`. Besides the already mentioned propositions $t_1, \ldots, t_n$ for tiles, we use $c_1, \ldots, c_{n'} {\equiv:} \boldsymbol{c}$ as an $n'$ bit counter ranging from 0 to $m-1$. Arithmetical operations involving this counter are described informally in quotes because these only plays a minor role. However, these operations have short encodings as CTL-formulas, that is, their size is polynomially bounded in $n'$. Additionally, the proposition `dir` is used to force two sons whenever $\alpha$ gets two.

Define $p^0{:=}\neg p$ and $p^1{:=}p$ for any proposition $p$. For a label $a \in \Sigma$ and a CTL-formula $\psi$, $!\mathtt{X}^a\psi{:=}\mathtt{EX}^a\mathtt{tt} \wedge \mathtt{AX}^a\psi$ denotes that there is at least one $a$-successor and $\psi$ hold at these successors. Moreover, instead of automata we also use regular expressions as annotations to CTL-formulas.

The tiling problem is translated into the formula

$$\varphi \;:=\; \text{“}\boldsymbol{c} = 0\text{”} \wedge \mathtt{AG}^{\{\varepsilon\} \,\cup\, \Sigma^* \{\mathtt{nextCol},\mathtt{nextRow}\}}\psi$$

where $\psi$ is the conjunction of the following lines and the automaton $\mathcal{A}$ is depicted in Fig. 3.

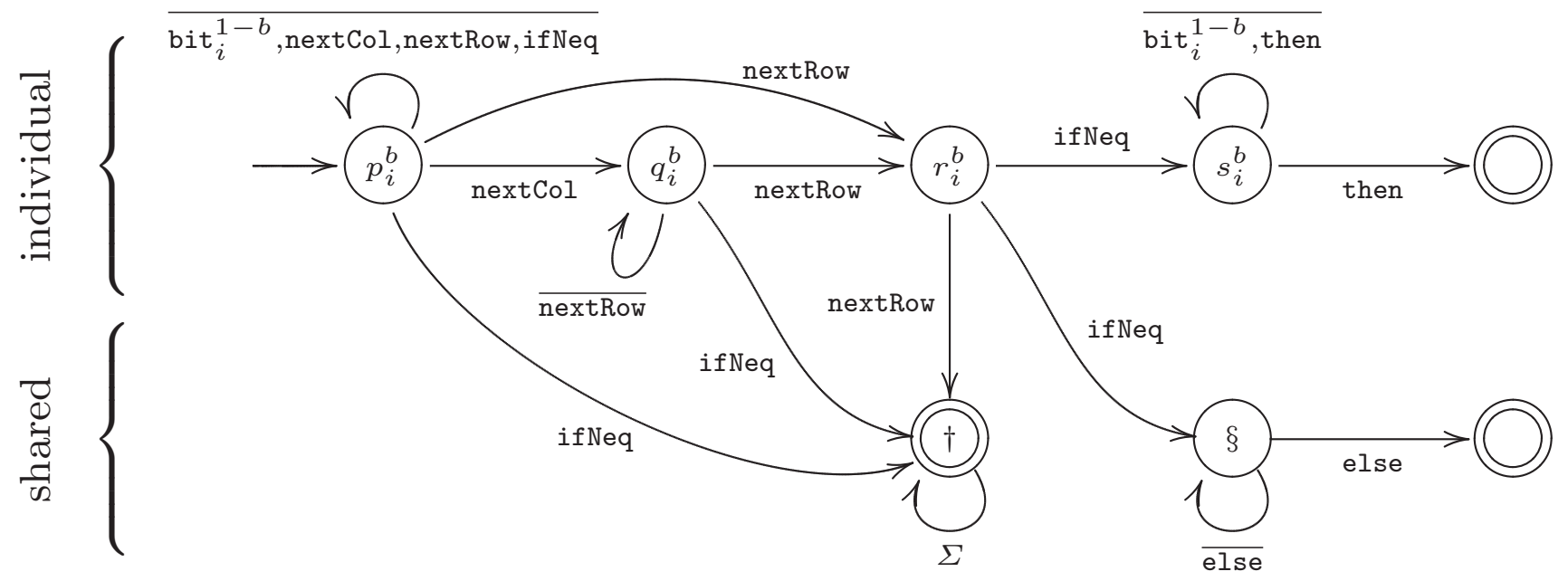

**Fig. 3.** Automaton $\mathcal{A}$. Overlined labels mean their complement with respect to $\Sigma$. The individual part is present for any $i \in [n]$ and for any $b \in \mathbb{B}$. So, it has $10n + 3$ states where $2n$ are initial ones.

$$\bigvee_{i \in [n]} t_i \wedge \bigwedge_{j \in [n] \setminus \{i\}} \neg t_j \tag{1}$$

$$\bigwedge_{i \in [n']} \bigwedge_{b \in \mathbb{B}} c_i^b \rightarrow \mathtt{AX}^{\Gamma^{i-1}} \mathtt{EX}^{\mathtt{bit}_i^b} \mathtt{tt} \tag{2}$$

$$\mathtt{EX}^{\mathtt{ifNeq}} \mathtt{tt} \tag{3}$$

$$\bigwedge_{i\in[n']} \bigwedge_{b\in\mathbb{B}} c_i^b \rightarrow \mathtt{AX}^{\mathtt{ifNeq}\ \Gamma^{i-1}} \mathtt{EX}^{\mathtt{bit}_i^b} \mathtt{tt} \tag{4}$$

$$\mathtt{AX}^{\mathtt{ifNeq}\ \Gamma^{n'}} !\mathtt{X}^{\mathtt{then}} \mathtt{dispose} \tag{5}$$

$$\mathtt{AX}^{\mathtt{ifNeq}\ \Gamma^{n'}\ \mathtt{then}} !\mathtt{X}^{\mathtt{else}} (\neg \mathtt{dispose} \wedge \mathtt{AX}\mathtt{ff}) \tag{6}$$

$$\bigwedge_{i\in[n]} (\mathtt{AX}^{\mathtt{ifNeq}\ \Gamma^{n'}\ \mathtt{then\ else}} t_i) \leftrightarrow t_i \tag{7}$$

$$\bigwedge_{i\in[n],\alpha(i)\neq 0} t_i \rightarrow \mathtt{AF}^{\mathcal{A}} \left( \bigvee_{j\in[n],iVj} t_j \vee \mathtt{dispose} \right) \tag{8}$$

$$\text{“}\boldsymbol{c} < m-1\text{”} \rightarrow \bigvee_{i,j\in[n],iHj} t_i \wedge \mathtt{AX}^{\Gamma^{n'}} !\mathtt{X}^{\mathtt{nextCol}} t_j \tag{9}$$

$$\text{“}\boldsymbol{c} < m-1\text{”} \rightarrow \text{“}\mathtt{AX}^{\Gamma^{n'}\ \mathtt{nextCol}} \boldsymbol{c} = \boldsymbol{c}+1\text{”} \tag{10}$$

$$\left( \text{“}\boldsymbol{c} = m-1\text{”} \wedge \bigvee_{i\in[n],\alpha(i)>0} t_i \right) \rightarrow \mathtt{AX}^{\Gamma^{n'}} !\mathtt{X}^{\mathtt{nextRow}} (\mathtt{dispose} \wedge \text{“}\boldsymbol{c}=0\text{”}) \tag{11}$$

$$\left( \text{“}\boldsymbol{c} = m-1\text{”} \wedge \bigvee_{i\in[n],\alpha(i)=2} t_i \right) \rightarrow \bigwedge_{b\in\mathbb{B}} \mathtt{AX}^{\Gamma^{n'}} \mathtt{EX}^{\mathtt{nextRow}} \mathtt{dir}^b \tag{12}$$

$$\left( \text{“}\boldsymbol{c} = m-1\text{”} \wedge \bigvee_{i\in[n],\alpha(i)=0} t_i \right) \rightarrow \mathtt{EX}^{\mathtt{ifNeq\ else}} \mathtt{dispose} \tag{13}$$

The formula $\varphi$ is obviously a CTL[DFA,NFA]-formula and its size is polynomially bounded in $n$.

The formula (1) ensures that exactly one tile is chosen, (2) programs the proof obligation (for the $V$-constraint) generated by (8). The verification is performed by (3)–(7). The formulas (9)–(12) ensure that the columns of a node in the tiling tree are enumerated, and that the tree is branching with respect to $\alpha$. The formula (13) is the counterpart to (9) and just ensures that proof obligation at the leaves are satisfied. (Alternatively, (2)–(7) could be excluded for the very last column.)

If we neglect the $V$-constraint, the reduction is sound and complete. As for the $V$-constraint, we describe the life of a proof obligation on a tree model of $\varphi$. An excerpt is given in Fig. 4.

Let $Q$ be the set of states of $\mathcal{A}$. If we say that there is a proof obligation in a certain state $Q' \subseteq Q$, we refer to the deterministic substitute of $\mathcal{A}$ obtained from the powerset construction. Beginning at the node 1, the formula (8) admits a proof obligation for $t_j \vee$ `dispose` (for some $j \in [n]$) in the state $\{p_i^b \mid i \in [n], b \in \mathbb{B}\}$. The intended trace is the first line in Fig. 4. After passing the label `nextRow` the automaton reaches the state $\{q_i^b \mid i \in [n], b \in \mathbb{B}, 1 \models c_i^b\}$, that is, the state reflect the content of the counter at node 1. As for the second line, the proof obligation vanishes because `dispose` holds at the node 6. Moreover, the obligation remains while passing another columns of the same row. Changing the row for the first time, the obligation changes to $\{r_i^b \mid i \in [n], b \in \mathbb{B}, 1 \models c_i^b\}$ where the node 1 refers to the node which admits the proof obligation. As long as we follow the first line, the state remains until we change the row for the second time. This brings the obligation in the state $\{\dagger\}$. The formulas 5 and 13 offers a node with models `dispose` and ensure that the proof obligation disappears. Note that after the first change of the row there is also a node modelling `dispose`. But the state of the obligation does not contain a final state of $\mathcal{A}$ at this time.

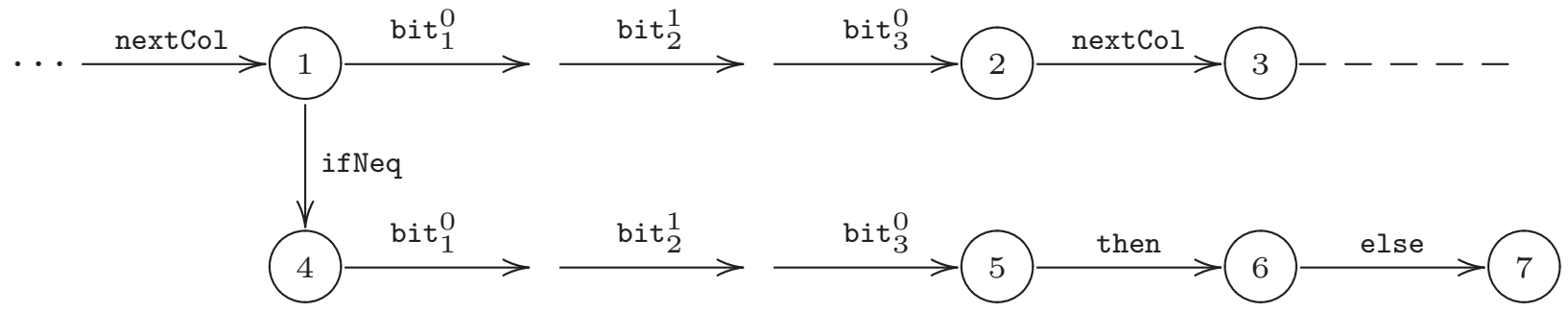

**Fig. 4.** Excerpt of a model for $\varphi$. This part depicts a single column which is neither the first nor the last one of a row. The second line shows the appendix which verifies the proof obligation for the $V$-constraint. At the node 1 the formulas $t_7$, $\neg c_3^1$, $c_2$ and $\neg c_1$ shall hold, at the node 6 the proposition `dispose`, and at node 7 the proposition $t_7$.

Now, we consider a proof obligation in the second line after passing `nextCol` for the first time. The label `ifNeq` switches the state to $\{\S, s_i^b \mid i \in [n], b \in \mathbb{B}, 1 \models c_i^b\}$. Again the node 1 refers to the node which admits the proof obligation. At node 5 the obligation either reaches the state $\{\S\}$ or some proper super set. The second case can only happen if the programmed counter and the counter of the current column differ. In this case, the formula (5) disposes the obligation. Otherwise, the state of the obligation does not contain a final state when reaching the node 6. By (6) and (7), the tile $t_j$—as represented by the obligation—must be the tile of the current column.

For the other parts involving DVPA, again, the constructed formula $\varphi$ shall imitate a successful tree of $T$ on the input. The space bound can be controlled by a counter with appropriate domain. The constraints between cells of consecutive configurations, however, are implemented differently. We use a deterministic VPA to push all cells along the whole branch of the run on the stack—configuration by configuration. At the end, we successively take the cells from the stack and branch. Along each branch, we use the counter to remove exponential or doubly exponential, resp., many elements from stack to access the cell at the same position in the previous configuration. So, as a main component of $\varphi$ we use either $\mathtt{AU}^{\mathcal{A}}\mathtt{AXff}$ or $\mathtt{AG}^{\mathcal{A}}\mathtt{ff}$ for some VPA $\mathcal{A}$. In the case of a counter with a doubly exponential domain, the technique explained for CTL[DFA, NFA] can be applied. But this time, a proof obligation expresses a bit number and its value. □

**Corollary 5.6.** *The lower bounds for the satisfiability problem are as in Fig. 2.*

*Proof.* As CTL is EXPTIME-hard [12], so is CTL[_, _]. The lower bound for PDL[DVPA], that is 2EXPTIME [23], is also a lower bound for CTL[DVPA,

_] due to Thm. 4.2. The picture is completed by Thm. 5.3 combined with Prop. 4.1(2). □

# 6 Model Checking

In this section we consider model-checking of CTL[$\mathfrak{A}$, $\mathfrak{B}$] against finite and infinite transition systems, represented by (visibly) pushdown systems.

## 6.1 Finite State Systems

The following table summarises the complexities of model checking CTL[$\mathfrak{A}$,$\mathfrak{B}$] in finite transition systems. Surprisingly, despite its greatly increased expressive power compared to CTL, CTL[PDA,DPDA] remains in P. In general, it is the class $\mathfrak{B}$ which determines the complexity. The table therefore only contains one row ($\mathfrak{A}$) and several columns ($\mathfrak{B}$). Note that PDA covers everything down to DFA while DPDA covers DVPA and DFA.

| | DPDA | NFA | VPA | PDA |
|---|---|---|---|---|
| PDA | P-complete | PSPACE-complete | EXP-complete | undecidable |

**Theorem 6.1.** *Model checking finite state systems against* CTL[PDA,DPDA] *is in P,* CTL[PDA,VPA] *is in EXPTIME, and* CTL[PDA,NFA] *is in PSPACE.*

*Proof.* To obtain a PTIME algorithm for CTL[PDA,DPDA] we observe that – as for plain CTL – we can model check a CTL[$\mathfrak{A}$,$\mathfrak{B}$] formula bottom-up for any $\mathfrak{A}$ and $\mathfrak{B}$. Starting with the atomic propositions one computes for all subformulas the set of satisfying states, then regards the subformula as a proposition. Hence, it suffices to give algorithms for $\mathtt{E}(x\mathtt{U}^{\mathcal{A}}y)$ and $\mathtt{E}(x\mathtt{R}^{\mathcal{B}}y)$ for propositions $x$ and $y$, which is done in the following two lemmas below.

That model-checking for CTL[PDA,VPA] is in EXPTIME then follows from the fact that every VPA can be translated into an equivalent DPDA, with potentially an exponential blow-up.

**Lemma 6.2.** *Model checking* $\mathtt{E}(x\mathtt{U}^{\mathcal{A}}y)$ *over a finite-state system for a PDA* $\mathcal{A}$ *is in P.*

*Proof.* We reduce the problem to the reachability problem for PDS. Let $\mathcal{T} = (\mathcal{S}, \rightarrow, \ell)$ be an LTS and $\mathcal{A} = (Q, \Sigma, \Gamma, \delta, q_0, F)$. We construct a PDS $\mathcal{A}_{\mathcal{T}} = (Q \times \mathcal{S}, \Gamma, \Delta, \ell)$, where $\Delta = \{((p, s), \gamma) \rightarrow ((q, t), w) \mid$ exists $a \in \Sigma$ s.t. $s \xrightarrow{a} t$ and $(p, a, \gamma) \rightarrow (q, w)$ and $x \in \ell(s)\}$.

Now consider the set of configurations $R = \{((p, s), w) \mid p \in F$ and $y \in \ell(s)$ and $w \in \Gamma^*\}$. This set is clearly regular and hence, using the algorithm of Bouajjani *et al* [6] we can compute the set of predecessor configurations $PC = \{((p, s), w) \in C(\mathcal{A}_{\mathcal{T}}) \mid ((p, s), w) \Rightarrow^* ((q, t), w') \in R\}$ in polynomial time.

We claim $s \models \mathtt{E}(x\mathtt{U}^{\mathcal{A}}y)$ iff $((q_0, s), \epsilon) \in PC$. Let $((q_0, s), \epsilon) \in PC$. Then $((q_0, s), \epsilon) \Rightarrow^* ((q, t), w')$ for some $((q, t), w') \in R$. But this means that there exists a sequence of configurations $c_0, \dots, c_n$, s.t. $c_0 = ((q_0, s), \epsilon)$ and $c_n = ((q,t), w')$ and $c_i \Rightarrow c_{i+1}$. This entails that there is a $w \in \Sigma^*$, s.t. $s \xrightarrow{w} s'$, along that path in every state $x$ holds (except for s') and $w$ is accepted by $\mathcal{A}$ since $q \in F$. Finally, since $t \in \ell(y)$, we have $s \models \mathtt{E}(x\mathtt{U}^{\mathcal{A}}y)$.

The other direction is similar. Note that by the construction in [6], the set $PC$ is regular and hence membership can be checked in polynomial time as well. □

**Lemma 6.3.** *Model checking* $\mathtt{E}(x\mathtt{R}^{\mathcal{B}}y)$ *over a finite-state system for a DPDA* $\mathcal{B}$ *is in P.*

*Proof.* The problem can be reduced to model checking a fixed LTL formula on a PDS. Let $\mathcal{T} = (\mathcal{S}, \rightarrow, \ell)$ be an LTS and $\mathcal{A} = (Q, \Sigma, \Gamma, \delta, q_0, F)$. We construct a PDS $\mathcal{A}_{\mathcal{T}} = (Q \times \mathcal{S} \cup \{g, b\}, \Gamma, \Delta, \ell')$, where $\ell'$ extends $\ell$ by $\ell'(b) = \{p_b\}$, for a fresh proposition $p_b$ and

$$((p,s),\gamma) \rightarrow \begin{cases} (g,\epsilon) & \text{if } s \in \ell'(x) \text{ and} \\ & (p \in F \text{ implies } s \in \ell'(y)) \\ \\ (b,\epsilon) & \text{if } p \in F \text{ and } s \notin \ell'(y) \\ \\ ((q,t),w) & \text{if none of the above match} \\ & \text{and there exists } a \in \Sigma, \text{ s.t.} \\ & s \xrightarrow{a} t \text{ and } (p,a,\gamma) \rightarrow (q,w) \\ & \text{for some } \gamma \in \Gamma, w \in \Gamma^* \end{cases}$$

Let $\mathcal{F}$ be the LTL formula $\mathtt{F}p_b$. We show that $s \models_{\mathcal{T}} \mathtt{E}(x\mathtt{R}^{\mathcal{A}}y)$ iff $((q_0,s),\epsilon) \not\models_{\mathcal{T}_{\mathcal{A}}} \mathcal{F}$. Assume $s \not\models \mathtt{E}(x\mathtt{R}^{\mathcal{A}}y)$. This means that on all paths starting in $s$, $(\neg x\mathtt{U}^{\mathcal{A}}\neg y)$ holds and hence on all paths a $w = a_1 \dots a_n \in L(\mathcal{A})$ and $s_0, \dots, s_n$ exist, s.t. $s_0 \xrightarrow{a_1} s_1 \dots \xrightarrow{a_n} s_n$, $s = s_0$ and for all $i \leq n$ we have $s_i \models \neg x$ and $s_n \models \neg y$.

It is clear that since $\mathcal{A}$ is deterministic, every path in the corresponding PDS (starting in $((q_0,s),\epsilon)$) labeled with such a $w$ runs through the state $((p,s_n),v)$, where $p \in F$ and $v \in \Gamma^*$. Since $y \notin \ell'(s_n)$, every such path ends in the next state which is $(b,\epsilon)$, where $p_b$ holds. Therefore the LTL formula $\mathcal{F}$ holds in $\mathcal{A}_{\mathcal{T}}$ with the initial state being $((q_0,s),\epsilon)$. The reverse direction is similar.

Finally, it is well known that model checking a fixed LTL formula on a PDS is in P [6]. Hence, model checking $\mathtt{E}(x\mathtt{R}^{\mathcal{A}}y)$ for a DPDA $\mathcal{A}$ is P-time. □

Finally, we can adapt Lemma 6.3 to obtain a PSPACE algorithm for CTL[PDA,NFA]. We reduce $\mathtt{E}(x\mathtt{R}^{\mathcal{B}}y)$ to the problem of checking a fixed LTL formula against a determinisation of the NFA $\mathcal{B}$. This can be further reduced to a repeated reachability problem over the product of a Büchi automaton and a determinisation of the NFA. Since we can determinise by a subset construction, we can use Savitch's algorithm [27] and an on-the-fly computation of the edge relation. Because Savitch's algorithm requires LOGSPACE over an exponential graph, the complete algorithm runs in PSPACE. This completes the proof of theorem 6.1. □

We now consider the lower bounds.

**Theorem 6.4.** *For fixed finite state transition systems of size 1, model checking for* EF[VPA] *is PTIME-hard,* EG[NFA] *is PSPACE-hard,* EG[VPA] *is EXPTIME-hard, and* EG[PDA] *is undecidable.*

We split the proof into separate lemmas.

**Lemma 6.5.** *For fixed transition finite-state systems of size 1, model checking for* EF[VPA] *is PTIME-hard.*

*Proof.* It is known that the emptiness problem for VPA is PTIME-hard. Now let $\mathcal{A}$ be a VPA and consider the transition system $\mathcal{T} = (\{s\}, \rightarrow, \ell)$ with $s \xrightarrow{a} s$ for every $a \in \Sigma$. Then $L(\mathcal{A}) \neq \emptyset$ iff $\mathcal{T}, s \models \mathtt{EF}^L\mathtt{tt}$. Note that $\mathcal{T}$ has one state but as many transitions as there are symbols in the underlying alphabet. Using a binary encoding of the alphabet reduces the number of transitions to two. □

**Lemma 6.6.** *For fixed finite-state transition systems of size 1, model checking for* EG[NFA] *is PSPACE-hard.*

*Proof.* The following problem, known as the $n$-tiling problem, is known to be PSPACE-hard [30]: the input is a unarily encoded number $n$, a finite set $T$ of elements called tiles, and two relations $H, V \subseteq T^2$. The tiling problem is then to decide whether or not there is a mapping $f : \mathbb{N} \times [n] \rightarrow T$, s.t. for all $i \in \mathbb{N}$ and all $j$ with $0 \leq j < n$ we have: $(f(i,j), f(i,j+1)) \in H$ and $(f(i,j), f(i+1,j)) \in V$. Such a mapping is called a valid tiling of the $n$-corridor.

Now, given such $n, T, H, V$, we construct a transition system $\mathcal{T}_T$ with one state $s$ only and a EG[NFA] formula $\varphi^{n,T}_{H,V}$ s.t. $\mathcal{T}_T, s \models \varphi^{n,T}_{H,V}$ iff a valid tiling of the $n$-corridor exists.

We use $\Sigma = T$ and $\mathcal{P} = \emptyset$. Let $\mathcal{T}_T = (\{s\}, \rightarrow, \ell)$ with $s \xrightarrow{t} s$ for all $t \in T$. Note that any path through $\mathcal{T}_{n,T}$ encodes in a straight-forward way a tiling of the $n$-corridor, listing the assigned tiles row-by-row. Furthermore, for every possible tiling of the $n$-corridor, there is a path in $\mathcal{T}_{n,T}$ starting in any state that represents this tiling.

Now we let $\varphi^{n,T}_{H,V}$ express that there is a tiling (i.e. a path) which is valid in the sense that no pair of tiles occurring in a position $i, i+1$ with $i \neq 0 \mod n$ is not in $H$, and no pair occurring in positions $i, i+n$ is not in $V$. Let $L :=$

$$\Big( \bigcup_{(t,t') \notin V} T^* \, t \, T^n \, t' \Big) \;\cup\; \Big( \bigcup_{(t,t') \notin H} (T^n)^* \, \big( \bigcup_{i=0}^{n-2} T^i \, t \, t' \big) \Big).$$

It is well-known that there is an NFA $\mathcal{A}$ with $L(\mathcal{A}) = L$ and $|\mathcal{A}| = \mathcal{O}(|T|^2 \cdot n^2)$. Hence, let $\varphi^{n,T}_{H,V} := \mathtt{EG}^{\mathcal{A}}\mathtt{ff}$. A binary encoding of the tiles yields a transition system of truely fixed size again. □

**Lemma 6.7.** *For fixed finite-state transition systems of size 1, model checking for* EG[VPA] *is EXPTIME-hard.*

*Proof.* Let $M = (\mathcal{Q}, \Gamma, \delta, \lambda)$ be a linear-space alternating Turing machine, where $\lambda$ is a function labelling each state in $\mathcal{Q}$ as existential, universal, accepting or rejecting. We assume, without loss of generality, that for each $q \in \mathcal{Q}$ and $a \in \Gamma$, $|\delta(q, a)| = 2$, and that the successor configurations are ordered into the first and second successor. There is some constant $k$ such that for every word $w$ of length $n$ accepted by $M$, $M$ uses at most $kn$ space. Furthermore, since $M$ is space-bounded, we can assume there are no infinite computations.

For a given word $w$, we define a transition system $\mathcal{T}$ with one state $s$ only and a EG[VPA] formula $\varphi_{M,w}$ s.t. $\mathcal{T}, s \models \varphi_{M,w}$ iff $w$ is accepted by $M$.

Let $\Sigma_M = \Gamma \cup (\Gamma \times \mathcal{Q})$. We use $\Sigma = \overleftarrow{\Sigma_M} \cup \overline{\Sigma_M} \cup \overrightarrow{\Sigma_M} \cup \delta \cup \{\overleftarrow{\#}, \overline{\#}, \overrightarrow{\#}, *\}$, where $\overleftarrow{\#}, \overline{\#}, \overrightarrow{\#}, * \notin \Gamma$, $\overline{\Sigma_M} = \{ \overline{a} \mid a \in \Sigma_M \}$ and similarly for $\overleftarrow{\Sigma_M}$ and $\overrightarrow{\Sigma_M}$, and $\mathcal{P} = \emptyset$. Let $\mathcal{T}_T = (\{s\}, \rightarrow, \ell)$ with $s \xrightarrow{t} s$ for all $t \in \Sigma$. $\mathcal{T}_T$ simply generates arbitrary infinite words.

We define $\varphi_{M,w} := \mathtt{EG}^{\mathcal{A}}\mathtt{ff}$ such that $\mathcal{A}$ accepts invalid or rejecting computations of $M$ over $w$. Configurations are encoded as words of the form $\Gamma^*(\Gamma \times \mathcal{Q})\Gamma^*$ and length $kn$. A computation of a space-bounded alternating Turing machine is a finite tree (since all computations are terminating). We encode a computation tree as a word, which traverses the tree in infix order. When progressing down the tree to an internal node $c$ we write $r\overrightarrow{\#c\#}$, where $r \in \delta$ is the rule applied (for convenience, we assume a null rule setting up the initial configuration), when reaching a leaf node we write $r\overline{\#c\#}$, and when backtracking to a node $c$ we write $\overleftarrow{\#c^R\#}$, where $\overrightarrow{c}$ is the configuration $c$ with each character $a$ replaced by $\overrightarrow{a}$, similarly for the other notations, and $c^R$ is the configuration $c$ reversed. Note that a node is only backtracked to once, from its first child. The second child backtracks to the node's parent. More precisely, let $leaf(c, r)$ denote a leaf node with configuration $c$ reached by rule $r$, and $int(t_l, c, t_r, r)$ denote a node with configuration $c$ reached by the rule $r$ and with the left sub-tree $l$ and right sub-tree $r$. Let

$$fl(leaf(c, r)) = r\overline{\#c\#}$$
$$fl(int(t_l, c, t_r, r)) = r\overrightarrow{\#c\#}fl(t_l)\overleftarrow{\#c^R\#}fl(t_r)\ .$$

Given a tree $T$, we flatten it to the word $fl(T)*^{\omega}$.

The VPA $\mathcal{A}$ is the union of $\mathcal{A}_1, \ldots, \mathcal{A}_9$ asserting the following properties.

1. The word is not of the form $d_0 \ldots d_m *^{\omega}$ where for all $i$, $d_i$ is of the form $\overleftarrow{\#c\#}$, $r\overline{\#c\#}$ or $r\overrightarrow{\#c\#}$, where $c$ is a valid configuration, all leaf configurations are followed by backtracks, and backtracks only occur after leaf configurations or backtrack configurations.
2. The word does not begin with the initial configuration.
3. The word does not encode a complete tree structure.
4. There is a leaf that is not an accepting configuration.
5. There is a $d_i = r\overrightarrow{\#c\#}$ encoding an existential configuration, and $d_{i+1}$ is not $r'\overrightarrow{\#c'\#}$ or $r'\overline{\#c'\#}$ for a successor configuration $c'$ derived from $c$ by $r'$. That is, an existential node has no valid child.

6. There is a $d_i = \overleftarrow{\#c\#}$ encoding an existential configuration, and $d_{i+1} = \overrightarrow{r\#c'\#}$ or $r\overline{\#c'\#}$ for some $c'$. That is, an existential node has more than one child.
7. There is a $d_i = r\overrightarrow{\#c\#}$ encoding a universal configuration $c$, and $d_{i+1}$ is not $r'\overrightarrow{\#c'\#}$ or $r'\overline{\#c'\#}$ where $c'$ is the first successor configuration derived by the rule $r'$. That is, the first child visited is not the first successor.
8. There is $d_i = \overleftarrow{\#c^R\#}$ encoding a universal configuration $c$, and $d_{i+1}$ is $\overleftarrow{\#c'\#}$. That is, after returning to a universal node, the second successor is not visited.
9. There is $d_i = \overleftarrow{\#c^R\#}$ encoding a universal configuration $c$, and $d_{i+1}$ is $\overrightarrow{r\#c'\#}$ or $r\overline{\#c'\#}$ but $c'$ is not the second successor configuration of $c$ derived by the rule $r$. That is, after returning to a universal node, and visiting a child, the child is not the second successor.

The automata $\mathcal{A}_1$, $\mathcal{A}_2$, $\mathcal{A}_4$, $\mathcal{A}_6$ and $\mathcal{A}_8$ are NFA that are straightforward to construct.

The automaton $\mathcal{A}_3$ is the complement of the deterministic VPA that pushes each character $\overrightarrow{a}$ it reads onto the stack, and pops each character $\overleftarrow{a}$, provided that $\overrightarrow{a}$ is the current top stack character. Leaf characters $\overline{a}$ and rules $r$ do not affect the stack. The $*$ symbol can only be seen when the stack is empty, and is the only character seen after its first appearance. The automaton accepts all words ending with a sequence of $*$. The complement of this automaton accepts all words that do not encode trees.

The automaton $\mathcal{A}_5$ is the union of the NFA $\mathcal{A}_r$ for each $r \in \delta$. Each NFA $\mathcal{A}_r$ accepts the union of the languages

1. $\Sigma^* \overrightarrow{\#\Gamma^*(a,q)\Gamma^*\#} r$ where $r$ is not applicable at (existential) state $q$ and head character $a$.
2. $\bigcup_{j=1}^{n} \Sigma^* \overrightarrow{\#\Gamma^{j-1}(a,q)\Gamma^*\#} r(\overrightarrow{w} \cup \overline{w})$ where $w = \#\Gamma^k\Gamma$ and $k = j-2$ if $r$ moves left, and $k = j$ if $r$ moves right, and $q$ is existential.
3. $\bigcup_{j=1}^{n} \Sigma^* \overrightarrow{\#\Gamma^{j-1}(a,q)\Gamma^*\#} r(\overrightarrow{w} \cup \overline{w})$ where $w = \#\Gamma^{j-1}b$ and $r$ does not write a $b$ character and $q$ is existential.
4. $\bigcup_{j=1}^{n} \Sigma^* \overrightarrow{\#\Sigma_M^{j-1} a \Sigma_M^* \#} r(\overrightarrow{w} \cup \overline{w})$ where $w = \#\Sigma_M^{j-1}(b \cup (b,q))$ with $a, b \in \Gamma$ and $a \neq b$ or $q$ is not the state moved to by $r$.

The automaton $\mathcal{A}_7$ is the union of the NFA $\mathcal{A}_r$ for each $r \in \delta$. Each NFA $\mathcal{A}_r$ accepts the union of the languages

1. $\Sigma^* \overrightarrow{\#\Gamma^*(a,q)\Gamma^*\#} r$ where $r$ is not applicable at (universal) state $q$ and head character $a$, or $r$ is not a first successor rule.
2. $\bigcup_{j=1}^{n} \Sigma^* \overrightarrow{\#\Gamma^{j-1}(a,q)\Gamma^*\#} r(\overrightarrow{w} \cup \overline{w})$ where $w = \#\Gamma^k\Gamma$ and $k = j-2$ if $r$ moves left, and $k = j$ if $r$ moves right, and $q$ is universal.
3. $\bigcup_{j=1}^{n} \Sigma^* \overrightarrow{\#\Gamma^{j-1}(a,q)\Gamma^*\#} r(\overrightarrow{w} \cup \overline{w})$ where $w = \#\Gamma^{j-1}b$ and $r$ does not write a $b$ character and $q$ is universal.

4. $\bigcup_{j=1}^{n} \Sigma^* \overrightarrow{\#\Sigma_M^{j-1} a \Sigma_M^* \#r}(\overrightarrow{w} \cup \overline{w})$ where $w = \#\Sigma_M^{j-1}(b \cup (b,q))$ with $a, b \in \Gamma$ and $a \neq b$ or $q$ is not the state moved to by $r$.

Finally $\mathcal{A}_9$ is the union of the NFA $\mathcal{A}_r$ for each rule $r \in \delta$. Each NFA $\mathcal{A}_r$ accepts the union of the languages

1. $\Sigma^* \overleftarrow{\#\Gamma^*(a,q)\Gamma^*\#r}$ where $r$ is not applicable at (universal) state $q$ and head character $a$, or $r$ is not a second successor rule.
2. $\bigcup_{j=1}^{n} \Sigma^* \overleftarrow{\#\Gamma^{j-1}(a,q)\Gamma^*\#r}(\overrightarrow{w} \cup \overline{w})$ where $w = \#\Gamma^k\Gamma$ and $k = n - j + 2$ if $r$ moves left, and $k = n - j$ if $r$ moves right, and $q$ is universal.
3. $\bigcup_{j=1}^{n} \Sigma^* \overleftarrow{\#\Gamma^{j-1}(a,q)\Gamma^*\#r}(\overrightarrow{w} \cup \overline{w})$ where $w = \#\Gamma^{n-j-1}b$ and $r$ does not write a $b$ character and $q$ is existential.
4. $\bigcup_{j=1}^{n} \Sigma^* \overleftarrow{\#\Sigma_M^{j-1} a \Sigma_M^* \#r}(\overrightarrow{w} \cup \overline{w})$ where $w = \#\Sigma_M^{n-j-1}(b \cup (b,q))$ with $a, b \in \Gamma$ and $a \neq b$ or $q$ is not the state moved to by $r$.

□

**Lemma 6.8.** *For fixed finite-state transition systems of size 1, model checking for* EG[PDA] *is undecidable.*

*Proof.* The proof is by a reduction from the octant tiling problem, which is known to be undecidable. Let $Oct = \{ (i,j) \mid 0 \leq j \leq i \}$, $T$ be a finite set of tiles, and $H, V \subset T^2$ be two relations. The problem is to find a mapping $F : Oct \to T$ such that, for all $(i,j) \in Oct$, $(f(i,j), f(i+1,j)) \in V$ and whenever $j < i$, $(f(i,j), f(i,j+1)) \in H$.

Given $T$, $H$ and $V$, we construct a transition system $\mathcal{T}_T$ with one state $s$ only and a EG[PDA] formula $\varphi_{H,V}^T$ s.t. $\mathcal{T}_T, s \models \varphi_{H,V}^T$ iff a valid tiling exists.

We use $\Sigma = T \cup \{\#\}$, where $\# \notin T$ and $\mathcal{P} = \emptyset$. Let $\mathcal{T}_T = (\{s\}, \to, \ell)$ with $s \xrightarrow{t} s$ for all $t \in \Sigma$. $\mathcal{T}_T$ simply generates arbitrary words. Let $\varphi_{H,V}^T := \mathtt{EG}^{\mathcal{A}}\mathtt{ff}$. We define $\mathcal{A}$ to accept all words that do not represent a valid tiling, encoded as a word $\#t_{1,1}\#t_{2,1}t_{2,2}\#t_{3,1}t_{3,2}t_{3,3}\#\cdots$, where $t_{i,j} = f(i,j)$.

The automaton $\mathcal{A}$ accepts the language $L = L_1 \cup L_\# \cup \cup L_H \cup L_V$ where,

$$\begin{aligned} L_1 &= \#T^2 \\ L_\# &= \Sigma^*\#T^n\#T^{n'}\# && \text{where } n' \neq n+1 \\ L_H &= \Sigma^* t t' && \text{where } (t,t') \notin H \\ L_V &= \Sigma^*\#T^n t T^* \#T^n t' && \text{where } (t,t') \notin V \end{aligned}$$

It is straightforward to encode $L$ as a pushdown system. Together $L_1$ and $L_\#$ cover the cases when the word does not encode an octant. The language $L_H$ finds violations of the matching $H$, and $L_V$ catches violations of $V$. Using a binary encoding of the tiles yields a transition system of fixed size. □

### 6.2 Visibly Pushdown Automata

We consider model checking over an infinite transition system represented by a visibly pushdown automaton. We have the following with undecidability for EF[DPDA].

| | DFA/ DVPA | NFA/ VPA | DPDA |
|---|---|---|---|
| DFA $\cdots$ VPA | EXPTIME-complete | 2EXPTIME-complete | undecidable |

**Theorem 6.9.** *Model checking VPA against* CTL[VPA,DVPA] *is in EXPTIME, and* CTL[VPA,VPA] *is in 2EXPTIME.*

We split the proof into separate lemmas. We write $(q, \gamma, a, push(b), q')$, $(q, \gamma, a, rew(b), q')$ and $(q, \gamma, a, pop, q')$ for VPA rules, and omit the input character $\gamma$ for PDS rules.

**Lemma 6.10.** *Model checking* CTL[VPA,DVPA] *over visibly pushdown automata is in EXPTIME.*

*Proof.* We reduce the model checking problem for CTL[VPA, DVPA] over VPA to a Büchi game over a PDS. Since deciding the winner in such a game is EXPTIME [34], we obtain an EXPTIME algorithm for the model checking problem.

Without loss of generality, we assume all VPA have a bottom of stack symbol that is neither popped nor pushed and are complete. We also assume all formulas are in positive normal form.

The game has the following transitions. The state set and alphabet is defined implicitly. We begin with some standard formula to game translation. The alphabet becomes a set of pairs, $(a, b)$. The first component corresponds to the model VPA, the second to the formula VPA being evaluated. All states annotated *begin* are controlled by the existential player. The universal positions are $(s, \varphi_1 \wedge \varphi_2)$. The following rules are for all characters $a$ and $b$.

- $(win, (a, b), rew((a, b)), win)$.
- $((s, p)^{begin}, (a, b), rew((a, b)), win)$ if $s$ satisfies the atomic proposition $p$.
- $((s, \neg p)^{begin}, (a, b), rew((a, b)), win)$ if $s$ does not satisfy the atomic proposition $p$.
- $((s, \varphi_1 \vee \varphi_2)^{begin}, (a, b), rew((a, b)), (s, \varphi_i)^{begin})$ for $i \in \{1, 2\}$.
- $((s, \varphi_1 \wedge \varphi_2)^{begin}, (a, b), rew((a, b)), (s, \varphi_1 \wedge \varphi_2))$.
- $((s, \varphi_1 \wedge \varphi_2), (a, b), rew((a, b)), (s, \varphi_i)^{begin})$ for $i \in \{1, 2\}$.

For path formulas, we form a product with the VPA labelling the formula. We begin by adding a bottom of stack symbol to the stack in the formula VPA's component. For $\mathtt{E}(\varphi_1 \mathtt{U}^A \varphi_2)$ we allow the existential player to decide whether to complete the until formula or postpone completion until later. When postponing, the opponent can check whether the until will eventually be completed, or whether the condition on the until holds. When progressing the game, the existential player is able to choose both the move of the formula VPA and the model VPA. The existential positions are $(s, \mathtt{E}(\varphi_1 \mathtt{U}^{A_q} \varphi_2))$ and $(s, \mathtt{E}(\varphi_1 \mathtt{U}^{A_q} \varphi_2), move)$. The universal positions are $(s, \mathtt{E}(\varphi_1 \mathtt{U}^{A_q} \varphi_2), wait)$.

- $((s, \mathtt{E}(\varphi_1 \mathtt{U}^A \varphi_2))^{begin}, (a, b), rew((a, \bot)), (s, \mathtt{E}(\varphi_1 \mathtt{U}^{A_{q_0^A}} \varphi_2)))$.
- $((s, \mathtt{E}(\varphi_1 \mathtt{U}^{A_q} \varphi_2)), (a, b), rew((a, b)), (s, \varphi_2)^{begin})$ for all $a, b$ and $q$ is accepting.
- $((s, \mathtt{E}(\varphi_1 \mathtt{U}^{A_q} \varphi_2)), (a, b), rew((a, b)), (s, \mathtt{E}(\varphi_1 \mathtt{U}^{A_q} \varphi_2), wait))$ for all $a, b$.
- $((s, \mathtt{E}(\varphi_1 \mathtt{U}^{A_q} \varphi_2), wait), (a, b), rew(a), (s, \varphi_1)^{begin})$ for all $a, b$.
- $((s, \mathtt{E}(\varphi_1 \mathtt{U}^{A_q} \varphi_2), wait), (a, b), rew((a, b)), (s, \mathtt{E}(\varphi_1 \mathtt{U}^{A_q} \varphi_2), move))$ for all $a, b$.

- $((s, \mathtt{E}(\varphi_1 \mathtt{U}^{A_q} \varphi_2), move), (a,b), push((a',b')), (s', \mathtt{E}(\varphi_1 \mathtt{U}^{A_{q'}} \varphi_2)))$ whenever we have the rules $(s, \gamma, a, push(a'), s')$ and $(q, \gamma, b, push(b'), q')$.
- $((s, \mathtt{E}(\varphi_1 \mathtt{U}^{A_q} \varphi_2), move), (a,b), rew((a',b')), (s', \mathtt{E}(\varphi_1 \mathtt{U}^{A_{q'}} \varphi_2)))$ whenever there is $(s, \gamma, a, rew(a'), s')$ and $(q, \gamma, b, rew(b'), q')$.
- $((s, \mathtt{E}(\varphi_1 \mathtt{U}^{A_q} \varphi_2), move), (a,b), pop, (s', \mathtt{E}(\varphi_1 \mathtt{U}^{A_{q'}} \varphi_2)))$ whenever $(s, \gamma, a, pop, s')$ and $(q, \gamma, b, pop, q')$.

The remaining path formulas are similar, but the roles of the players are altered accordingly. In the case $\mathtt{A}(\varphi_1 \mathtt{U}^{A} \varphi_2)$, when satisfaction is postponed, since the property must hold for all paths, first the opponent picks a transition of the model, then the existential player picks a move in $A$. The existential positions are $(s, \mathtt{A}(\varphi_1 \mathtt{U}^{A_q} \varphi_2))$ and $(s, \mathtt{A}(\varphi_1 \mathtt{U}^{A_q} \varphi_2), t_s)$. The universal positions are $(s, \mathtt{E}(\varphi_1 \mathtt{U}^{A_q} \varphi_2), wait)$. Note that $\mathtt{A}(\varphi_1 \mathtt{U}^{A} \varphi_2)$ is an abbreviation for a $\neg \mathtt{E}(\neg \varphi_1 \mathtt{R}^{A} \neg \varphi_2)$. Due to the discussion in Section 2, correctness of the reduction relies on $A$ being deterministic.

- $((s, \mathtt{A}(\varphi_1 \mathtt{U}^{A} \varphi_2))^{begin}, (a,b), rew((a,\bot)), (s, \mathtt{A}(\varphi_1 \mathtt{U}^{A_{q_0^A}} \varphi_2)))$.
- $((s, \mathtt{A}(\varphi_1 \mathtt{U}^{A_q} \varphi_2)), (a,b), rew((a,b)), (s, \varphi_2)^{begin})$ and $q$ is accepting.
- $((s, \mathtt{A}(\varphi_1 \mathtt{U}^{A_q} \varphi_2)), (a,b), rew((a,b)), (s, \mathtt{A}(\varphi_1 \mathtt{U}^{A_q} \varphi_2), wait))$.
- $((s, \mathtt{A}(\varphi_1 \mathtt{U}^{A_q} \varphi_2), wait), (a,b), rew((a,b)), (s, \varphi_1)^{begin})$.
- $((s, \mathtt{A}(\varphi_1 \mathtt{U}^{A_q} \varphi_2), wait), (a,b), rew((a,b)), (s, \mathtt{A}(\varphi_1 \mathtt{U}^{A_q} \varphi_2), t_s))$ where $t_s$ is a transition from $s, a$.
- $((s, \mathtt{A}(\varphi_1 \mathtt{U}^{A_q} \varphi_2), t_s), (a,b), push((a',b')), (s', \mathtt{A}(\varphi_1 \mathtt{U}^{A_{q'}} \varphi_2)))$ whenever we have $t_s = (s, \gamma, a, push(a'), s')$ and $(q, \gamma, b, push(b'), q')$.
- $((s, \mathtt{A}(\varphi_1 \mathtt{U}^{A_q} \varphi_2), t_s), (a,b), rew((a',b')), (s', \mathtt{A}(\varphi_1 \mathtt{U}^{A_{q'}} \varphi_2)))$ whenever we have $t_s = (s, \gamma, a, rew(a'), s')$ and $(q, \gamma, b, rew(b'), q')$.
- $((s, \mathtt{A}(\varphi_1 \mathtt{U}^{A_q} \varphi_2), t_s), (a,b), pop, (s', \mathtt{A}(\varphi_1 \mathtt{U}^{A_{q'}} \varphi_2)))$ whenever $t_s = (s, \gamma, a, pop, s')$ and $(q, \gamma, b, pop, q')$.

The release operators are defined analogously. We begin with $\mathtt{E}(\varphi_1 \mathtt{R}^{A} \varphi_2)$. The existential positions are $(s, \mathtt{E}(\varphi_1 \mathtt{R}^{A_q} \varphi_2))$ and $(s, \mathtt{E}(\varphi_1 \mathtt{R}^{A_q} \varphi_2), move)$. The universal positions are $(s, \mathtt{E}(\varphi_1 \mathtt{R}^{A_q} \varphi_2), wait)$ and $(s, \mathtt{E}(\varphi_1 \mathtt{R}^{A_q} \varphi_2), t_s)$. Here we also rely on the fact that the VPA in the formulas are deterministic.

- $((s, \mathtt{E}(\varphi_1 \mathtt{R}^{A} \varphi_2))^{begin}, (a,b), rew((a,\bot)), (s, \mathtt{E}(\varphi_1 \mathtt{R}^{A_{q_0^A}} \varphi_2)))$.
- $((s, \mathtt{E}(\varphi_1 \mathtt{R}^{A_q} \varphi_2)), (a,b), rew((a,b)), (s, \varphi_1)^{begin})$.
- $((s, \mathtt{E}(\varphi_1 \mathtt{R}^{A_q} \varphi_2)), (a,b), rew((a,b)), (s, \mathtt{E}(\varphi_1 \mathtt{R}^{A_q} \varphi_2), wait))$.
- $((s, \mathtt{E}(\varphi_1 \mathtt{R}^{A_q} \varphi_2), wait), (a,b), rew((a,b)), (s, \varphi_1)^{begin})$ where $q$ is accepting.
- $((s, \mathtt{E}(\varphi_1 \mathtt{R}^{A_q} \varphi_2), wait), (a,b), rew((a,b)), (s, \mathtt{A}(\varphi_1 \mathtt{U}^{A_q} \varphi_2), move))$.
- $((s, \mathtt{E}(\varphi_1 \mathtt{R}^{A_q} \varphi_2), move), (a,b), rew((a,b)), (s, \mathtt{A}(\varphi_1 \mathtt{U}^{A_q} \varphi_2), t_s))$ where $t_s$ is a transition from $s, a$.
- $((s, \mathtt{E}(\varphi_1 \mathtt{R}^{A_q} \varphi_2), t_s), (a,b), push((a',b')), (s', \mathtt{E}(\varphi_1 \mathtt{R}^{A_{q'}} \varphi_2)))$ whenever we have $t_s = (s, \gamma, a, push(a'), s')$ and $(q, \gamma, b, push(b'), q')$.
- $((s, \mathtt{E}(\varphi_1 \mathtt{R}^{A_q} \varphi_2), t_s), (a,b), rew((a',b')), (s', \mathtt{E}(\varphi_1 \mathtt{R}^{A_{q'}} \varphi_2)))$ whenever we have $t_s = (s, \gamma, a, rew(a'), s')$ and $(q, \gamma, b, rew(b'), q')$.
- $((s, \mathtt{E}(\varphi_1 \mathtt{R}^{A_q} \varphi_2), t_s), (a,b), pop, (s', \mathtt{E}(\varphi_1 \mathtt{R}^{A_{q'}} \varphi_2)))$ whenever $t_s = (s, \gamma, a, pop, s')$ $(q, \gamma, b, pop, q')$.

And finally, $\mathtt{A}(\varphi_1 \mathtt{R}^A \varphi_2)$. The existential positions are $(s, \mathtt{A}(\varphi_1 \mathtt{R}^{A_q} \varphi_2))$. The universal positions are $(s, \mathtt{E}(\varphi_1 \mathtt{R}^{A_q} \varphi_2), \mathit{wait})$.

- $((s, \mathtt{A}(\varphi_1 \mathtt{R}^A \varphi_2))^{begin}, (a, b), \mathit{rew}((a, \bot)), (s, \mathtt{A}(\varphi_1 \mathtt{R}^{A_{q_0^A}} \varphi_2)))$.
- $((s, \mathtt{A}(\varphi_1 \mathtt{R}^{A_q} \varphi_2)), (a, b), \mathit{rew}((a, b)), (s, \varphi_1)^{begin})$.
- $((s, \mathtt{A}(\varphi_1 \mathtt{R}^{A_q} \varphi_2)), (a, b), \mathit{rew}((a, b)), (s, \mathtt{A}(\varphi_1 \mathtt{R}^{A_q} \varphi_2), \mathit{wait}))$.
- $((s, \mathtt{A}(\varphi_1 \mathtt{R}^{A_q} \varphi_2), \mathit{wait}), (a, b), \mathit{rew}(a), (s, \varphi_2)^{begin})$ where $q$ is accepting.
- $((s, \mathtt{A}(\varphi_1 \mathtt{R}^{A_q} \varphi_2), \mathit{wait}), (a, b), \mathit{push}((a', b')), (s', \mathtt{A}(\varphi_1 \mathtt{R}^{A_{q'}} \varphi_2)))$ whenever we have $(s, \gamma, a, \mathit{push}(a'), s')$ and $(q, \gamma, b, \mathit{push}(b'), q')$.
- $((s, \mathtt{A}(\varphi_1 \mathtt{R}^{A_q} \varphi_2), \mathit{wait}), (a, b), \mathit{rew}((a', b')), (s', \mathtt{A}(\varphi_1 \mathtt{R}^{A_{q'}} \varphi_2)))$ whenever we have $(s, \gamma, a, \mathit{rew}(a'), s')$ and $(q, \gamma, b, \mathit{rew}(b'), q')$.
- $((s, \mathtt{A}(\varphi_1 \mathtt{R}^{A_q} \varphi_2), \mathit{wait}), (a, b), \mathit{pop}, (s', \mathtt{A}(\varphi_1 \mathtt{R}^{A_{q'}} \varphi_2)))$ whenever $(s, \gamma, a, \mathit{pop}, s')$ and $(q, \gamma, b, \mathit{pop}, q')$.

The game has a Büchi winning condition. All states are accepting except for states containing an $\mathtt{U}$ operator. Since these formulas must always eventually be satisfied, they are not accepting. Since we assume all VPA are complete, play will only get stuck when a literal is not satisfied, in which case the existential player will lose.

Given a CTL[VPA] formula $\varphi$ and a VPA $B$, we can check whether $B$ satisfies $\varphi$ by asking whether the existential player wins the game described above from the control state $(s_0, \varphi^{begin})$ with the initial stack contents. Such games can be solved in EXPTIME [34]. □

**Lemma 6.11.** *Model checking* CTL[VPA,VPA] *over visibly pushdown automata is in 2EXPTIME.*

*Proof.* The proof follows from the exponential cost of determinising the VPA, and Lemma 6.10. □

**Theorem 6.12.** *Model checking VPA against* CTL[DFA] *is EXPTIME-hard,* EG[NFA] *is hard for 2EXPTIME, and* EF[DPDA] *and* EG[DPDA] *are undecidable.*

The theorem is proven in several lemmas.

**Lemma 6.13.** *Model checking* EG[NFA] *over visibly pushdown automata is 2EXPTIME-hard.*

*Proof.* We obtain 2EXPTIME-hardness by adapting Bozzelli's proof of the 2EXPTIME-hardness of CTL over PDS [22].

We reduce from an alternating TM with an exponential space work tape and one-way input tape. The VPA we construct manages the alternation by using the stack to perform a depth first search of the computation tree. The main difficulty is encoding an exponential space bounded tape with a polynomial formula and VPA. The trick is to output an arbitrary length tape, and use the power of CTL[REG] to reject any tape configurations that are not exponential.

For convenience we will construct a PDS rather than a VPA. To obtain a VPA, firstly we must add output symbols. Hence, we simply output a tuple containing the control state and top of stack symbol that has just been left. We also assume the output symbol contains the propositions from $\{b, f, l, op, check\}$ (defined below) that the last configuration satisfied. This is now a PDA, but not a VPA. However, this can easily be converted into a VPA by marking the output symbols appropriately. It is then straightforward to adjust the CTL[REG] formula to be insensitive to this marking, with only a linear blow up. Intuitively, each occurrence of a letter $a$ in a regular expression, will be replaced with $(a_{push} \cup a_{pop} \cup a_{rew})$.

We write $(p, a, push_w, p')$ for pushdown rules: when the current control state is $p$, and top of stack $a$, move to $p'$ and replace $a$ with the word $w$.

Let $\Sigma$ be the tape alphabet of the TM and $\mathcal{Q}$ be the set of control states. The tape configurations are represented as words of the form

$$bin_n(0)c_0 bin_n(1)c_1 \cdots bin_n(2^n - 1)c_{2^n-1}$$

where $bin_n(i)$ is the $n$-digit binary representation of $i$ with the least significant bit first. It is beyond the power, however, of a polynomially sized VPA to output only sequences that count correctly from 0 to $2^n - 1$. Hence, the VPA generates, nondeterministically, a word of the form

$$(\{0,1\}^n (\Sigma \cup (\Sigma \times \mathcal{Q})))^* \ .$$

One can check whether a configuration of this form is of the correct length as follows: first assert that the first $n$-digit binary number is $0^n$. Then check that every pair $b_i c_i b_{i+1} c_{i+1}$ has $b_i = b_{i+1} + 1$. Finally, the last number must be all 1s.

Execution proceeds as follows. We first write the initial configuration to the stack. We then guess the next configuration. Afterwards we have a branch. Either we can continue the execution, or we can check the last configuration against the previous configuration. The execution branches are marked *op*, and the check branch marked check. The CTL[REG] formula takes the shape

$$E(op \wedge AX(check \Rightarrow \varphi_{good}) U fin) \ .$$

That is, we look for an execution branch that passes all checks on the way. To implement the alternation, we back track by popping from the stack when an accepting path has been found. Upon returning to a universal configuration, we continue the execution by exploring the next successor configuration until all possibilities have be tried.

The checking phase of branch proceeds as follows. One branch simply pops the configuration from the stack. This branch allows the CTL[REG] formula to check the length of the configuration.

The remaining branches remove the tape configuration from the stack, but nondeterministically mark one position with the proposition $check_1$ (hence there are a number of branches for this step). Then the PDA moves to the previous configuration and does the same as before: pops the tape from the stack, nondeterministically marking a position with $check_2$. The CTL[REG] formula can test consistency by checking whether the two positions correspond to the same position in the work tape. If so, the markers $check_1$ and $check_2$ can be used to test the contents of the cell (and it's neighbours) to ensure a valid update has taken place.

More formally, given a word $w$ of length $n$ and an alternating TM $\mathcal{T}$ augmented with an $2^n$-space bounded two-way work-tape, we define the pushdown automaton $\mathcal{P}^w_{\mathcal{T}}$. We assume without loss of generality that two rules can be applied from each configuration, and which can be referred to as the first and second rules respectively. We also assume the initial state is existential.

To aid notation, we will write $(p, a, o, p')$ for a pushdown rule moving from control state $p$ to $p'$, when $a$ is on top of the stack, and $o$ is $push_w$ or $pop$. Our push command replaces $a$ with the word $w$. Sometimes $o$ will be a sequence of commands that can be easily simulated using intermediate states.

We define $\mathcal{P}^w_{\mathcal{T}}$ to have the following transitions. The set of states is defined implicitly. The initial state is $init$. To initialise the automaton we have

– $(init, \bot, (push_{fin\bot}; push_w), (continue, 0, p_0, \Box))$ where $w = E0^n(\Box, p_0)(\{0,1\}^n \Box)^k\#$ for any arbitrary $k$.

The main loop of the simulation is given by the *continue* states, which have the following rules. We store the current word position $i$, TM state $p$ and read cell character $a$ on the stack (to aid backtracking) before continuing the execution. We the simulate a move by guessing the next configuration. The branch phase is where the automaton can either check the consistency of the guessed tape, or continue the execution.

– $((continue, i, p, a), done, push_{done}, (back, i))$
– $((continue, i, p, a), x, push_{(i,p,a)x}, (move, i, x, p, a))$ for all $a \in \Sigma$ and $p \in \mathcal{Q}$ and $x \in \{E, A_1, A_2\}$ and $p$ is not accepting or $i \neq |w| + 1$.
– $((continue, i, p, a), x, pop_1, (back, i))$ for all $a \in \Sigma$ and $p \in \mathcal{Q}$ and $x \in \{E, A_1, A_2\}$ and $p$ is accepting and $i = |w| + 1$.
– $((move, i, x, p, a), \gamma, push_{yw}, (branch, r, i, p', a'))$ for all $a \in \Sigma$, $p \in \mathcal{Q}$, $x \in \{E, A_1, A_2\}$, and some $\gamma$. Furthermore when $x = E$, $r$ is any rule applicable at $p, a, \gamma, w(i)$; when $x = A_1$, $r$ is the first transition for $p, a, \gamma, w(i)$; and when $x = A_2$, $r$ is the second transition for $p, a, \gamma, w(i)$. In all cases, $r$ leads to state $p'$. The character $a'$ is the cell contents marked as the current head positions and $y$ is $E$ when $r$ moves to an existential state, and $A_1$ when moving to a universal state. Finally, $w$ is any word of the form

$$(\{0,1\}^n (\Sigma \cup (\Sigma \times \mathcal{Q})))^*$$

containing a single pair $(a', p')$. Note, we do not have an exponential number of rules, one for each $w$, but rather use polynomial intermediate states, from which appropriate next characters to push can be chosen.
– $((branch, r, i, p, a), \gamma, push_\gamma, (continue, j, p, a))$ where $j = i$ if $r$ is an $\epsilon$-transition in the input, or $j = i+1$ otherwise; and $((branch, r, i, p, a), \gamma, push_\gamma, (check, r))$ for all $r$ and $\gamma$.

The back phase implements the backtracking, and requires the following rules.

- $((back, i), a, pop_1, (back, i))$ for all $a \neq \#$.
- $((back, i), \#, pop_1, (back_1, i))$ for all $a \neq \#$.
- $((back_1, i), (j, p, a), pop_1, (back_2, j, p, a))$.
- $((back_2, i, p, a), E, push_{done}, (continue, i, p, a))$.
- $((back_2, i, p, a), A_1, push_{A_2}, (continue, i, p, a))$.
- $((back_2, i, p, a), A_2, push_{done}, (continue, i, p, a))$.

Finally, the check phase is as described above.

- $((check, r), x, pop_1, check_{len})$ for all $x \in \{E, A1, A2\}$.
- $((check, r), x, pop_1, check_{cells})$ for all $x \in \{E, A1, A2\}$.
- $(check_{len}, a, pop_1, check_{len})$ for all $a \neq \#$ and $(check_{len}, \#, push_{\#}, done)$.
- $((check_{cells}, r), a, pop_1, (check_{cells}, r))$ for all $a \neq \#$.
- $((check_{cells}, r), a, pop_1, (check_1, r))$ for all $a \neq \#$.
- $((check_1, r), a, pop_1, (check^1_{cells}, r))$ for all $a \neq \#$.
- $((check^1_{cells}, r), a, pop_1, (check^1_{cells}, r))$ for all $a \neq \#$.
- $((check^1_{cells}, r), \#, (pop_1)^3, (check^2_{cells}, r))$.
- $((check^2_{cells}, r), a, pop_1, (check^2_{cells}, r))$ for all $a \neq \#$.
- $((check^2_{cells}, r), a, pop_1, (check_2, r))$ for all $a \neq \#$.
- $((check_2, r), a, pop_1, (check^3_{cells}, r))$ for all $a \neq \#$.
- $((check^3_{cells}, r), a, pop_1, (check^3_{cells}, r))$ for all $a \neq \#$.
- $((check^3_{cells}, r), \#, push_{\#}, done))$.

The atomic proposition $op$ is true during the init, continue, clear, move, branch and back phases. The proposition $check$ is true during the check phase, with $check_{len}$, $check_{cells}$, $check_1$ and $check_2$ true at their respective control states, $done$ true at state $done$ and $fin$ at state $fin$. Furthermore, there is a proposition $a$ for each stack character $a$, which is true whenever $a$ is on the top of the stack. For binary digits, we have 1 which is true when 1 is on top of the stack and similarly for 0. Finally $b$ is true at the beginning of each block of $n$ binary digits, $f$ is true at the first block of a configuration, and $l$ at the last.

The corresponding CTL[REG] formula is defined below. The main part of the formula is identical to Bozzelli's. The change is the use of the `EG` operator towards the end. For convenience, for every proposition or control state or stack alphabet component $a$, we have a regular expression $[a]$ which is the union of all output tuples containing the proposition $a$. Call this alphabet $\Gamma$.

For any $n$, we define

$$\varphi_n = E\left((op \wedge AX(check \rightarrow (\varphi_{len} \wedge \varphi_{cells}))\, U\, fin\right)$$

where

$$\varphi_{len} = AX(check_{len} \rightarrow \varphi'_{len})$$

$$\varphi'_{len} = AG\left( \begin{array}{l} \left((b \wedge f) \rightarrow \bigwedge_{j=0}^{n-1} (AX)^j 0\right) \wedge \left((b \wedge l) \rightarrow \bigwedge_{j=0}^{n-1} (AX)^j 1\right) \wedge \\ \begin{bmatrix} (b \wedge \neg f) \rightarrow \\ \bigvee_j^{n-1} \begin{bmatrix} (AX)^j (0 \wedge (AX)^{n+2} 1) \wedge \\ \bigwedge_{i>j} (AX)^i (1 \wedge (AX)^{n+2} 0) \wedge \\ \bigwedge_{i<j} (AX)^i (1 \iff (AX)^{n+2} 1) \end{bmatrix} \end{bmatrix} \end{array} \right).$$

and $\varphi_{cells}$ is as described below. We look $(n+2)$ steps ahead since there are three steps between the inspection of each tape: $(pop_1)^3$. First we insist that on every $check_{cells}$ branch, either $check_2$ is never defined (which catches the case when there is no predecessor configuration), or, after every $check_1$ there is some path where $check_2$ is placed correctly, and the tape is consistent. Again, the formula is very similar to Bozzelli's, except we now need to use CTL[REG] rather than CTL*.

$$\begin{aligned}\varphi_{cells} &= AX(check_{cells} \rightarrow \varphi'_{cells})\\ \varphi'_{cells} &= AG(\neg check_2) \vee AG((check_1 \wedge b) \rightarrow \mathtt{EG}^{\mathcal{L}}\mathtt{ff})\end{aligned}$$

where $\mathtt{EG}^{\mathcal{L}}\mathtt{ff}$ looks for violations of either the positioning of $check_2$ (which must be at the corresponding location in the tape to $check_1$), or the consistency of the tape. If there exists a path with no violations, the formula is satisfied and the check is passed. Note that the choice of paths only depends on the placement of the checks. We have

$$\mathcal{L} = \mathcal{L}_{pos} \cup \mathcal{L}_{tape}$$

where $[check_2^b] = [check_2] \cap [b]$ and

$$\begin{aligned}\mathcal{L}_{pos} &= \textstyle\bigcup_{j=0}^{n-1} \mathcal{L}_{pos}^j\\ \mathcal{L}_{pos}^j &= \left(\Gamma^j[0]\Gamma^*[check_2^b]\Gamma^j[1]\right) \cup \left(\Gamma^j[1]\Gamma^*[check_2^b]\Gamma^j[0]\right)\end{aligned}$$

and to define $\mathcal{L}_{tape}$, we use, for shorthand, functions $Next_r(\sigma_1, \sigma_2, \sigma_3) = \sigma_4$ that specify the contents of the current ($j$th) cell $d$ with respect to the rule fired $r$ and the contents of the previous configuration's $(j-1)th$, $j$th and $(j+1)th$ cells $\sigma_1, \sigma_2$ and $\sigma_3$ respectively. Let $\#_l$ and $\#_r$ denote, for the benefit of $Next_r(\sigma_1, \sigma_2, \sigma_3)$, the left and right boundaries of the tape respectively.

$$\mathcal{L}_{tape} = \bigcup_r \left(r \cap \left(\left([f] \cap \mathcal{L}_f^r\right) \cup \left(\overline{[f] \cup [l]} \cap \mathcal{L}_i^r\right) \cup \left([l] \cap \mathcal{L}_l^r\right)\right)\right)$$

where

$$\begin{aligned}\mathcal{L}_f^r &= \textstyle\bigcup_{Next_r(\#_l, \sigma_1, \sigma_2) \neq \sigma_3} [\sigma_3]\Gamma^*[check_2^b]\Gamma^n[\sigma_1]\Gamma^{n+1}[\sigma_2]\\ \mathcal{L}_i^r &= \textstyle\bigcup_{Next_r(\sigma_1, \sigma_2, \sigma_3) \neq \sigma_4} [\sigma_4]\Gamma^*[\sigma_1][check_2^b]\Gamma^n[\sigma_2]\Gamma^{n+1}[\sigma_3]\\ \mathcal{L}_l^r &= \textstyle\bigcup_{Next_r(\sigma_1, \sigma_2, \#_r) \neq \sigma_3} [\sigma_3]\Gamma^*[\sigma_1][check_2^b]\Gamma^n[\sigma_2]\end{aligned}$$

A word $w$ is accepted by $\mathcal{T}$ iff $\mathcal{P}_{\mathcal{T}}^w$ satisfies the formula $\varphi_{|w|}$. Furthermore, $\mathcal{P}_{\mathcal{T}}^w$ and $\varphi_{|w|}$ are polynomial in the size of $w$ and $\mathcal{T}$. This completes the proof. □

**Lemma 6.14.** *Model checking* EF[DPDA] *against visibly pushdown automata is undecidable.*

*Proof.* We encode a two-counter machine, whose reachability problem is known to be undecidable. We write $q \xrightarrow[o_1]{o_2} q'$ to indicate a transition from control state $q$ to $q'$ while performing $o_i \in \left\{inc_i, dec_i, zero_i^?\right\}$ on the $i$th counter, as usual. We test reachability of a designated control state $q_f$ from an initial state $q_0$.

We first construct a PDA which encodes the machine and the first counter. It outputs the operations on the second counter. We then have $\mathtt{EF}^A\mathtt{tt}$, where $A$ tests for consistency of the second counter.

More formally, our pushdown system has the following rules. The characters $o_2^c$ are operations on the second counter, marked as call, return or internal. This makes the model a VPA.

- $(q, o_2^c, x, push(1), q')$ where $x \in \{0,1\}$ and $q \xrightarrow[inc_1]{o_2} q'$.
- $(q, o_2^r, 1, pop, q')$ where $q \xrightarrow[dec_1]{o_2} q'$.
- $(q, o_2^i, 0, rew(0), q')$ where $q \xrightarrow[zero_1^?]{o_2} q'$.
- $(q_f, fin, x, rew(x), q_{done})$ where $x \in \{0,1\}$.

Let $y$ range over $\{c, r, i\}$ and $x$ over $\{0,1\}$, then $A$ has the rules: $(q, (inc_2)^y, x, push(1), q)$, $(q, (dec_2)^y, 1, pop, q)$, $(q, (zero_2^?)^y, 0, rew(0), q)$, and $(q, fin, x, rew(x), q')$. The initial control is $q$ and stack is 0. Similarly for the model VPA. The state $q'$ is accepting.

It is easy to see that the counter machine can reach $q_f$ iff the configuration $q_0 0$ satisfies $\texttt{EF}^A\texttt{tt}$. □

**Lemma 6.15.** *Model checking* EG[DPDA] *over visibly pushdown systems is undecidable.*

*Proof.* As above, we encode a two-counter machine. However, instead of reducing reachability, we ask whether the machine has an infinite computation. This is also undecidable. We use the formula $\texttt{EG}^A\texttt{ff}$ where $A$ accepts violations in the consistency of the second counter, or *quit* moves. That is, the automaton must have an infinite run that never cheats. We ensure all maximal paths are infinite by allowing the automaton to give up at any point, and loop in a (losing) sink state.

Our pushdown system has the following rules. The characters $o_2^c$ are operations on the second counter, marked as call, return or internal. This makes the model a VPA.

- $(q, o_2^c, x, push(1), q')$ where $x \in \{0,1\}$ and $q \xrightarrow[inc_1]{o_2} q'$.
- $(q, o_2^r, 1, pop, q')$ where $q \xrightarrow[dec_1]{o_2} q'$.
- $(q, o_2^i, 0, rew(0), q')$ where $q \xrightarrow[zero_1^?]{o_2} q'$.
- $(q, quit, x, rew(x), q_{quit})$ where $x \in \{0,1\}$ and $q$ is a state of the counter machine, or $q_{quit}$.

Let $y$ range over $\{c, r, i\}$ and $x$ over $\{0,1\}$. The automaton $A$ has the rules: $(q, (inc_2)^y, x, push(1), q)$, $(q, (dec_2)^y, 1, pop, q)$, $(q, (dec_2)^y, 0, pop, q')$, $(q, (zero_2^?)^y, 0, rew(0), q)$, $(q, (zero_2^?)^y, 1, rew(1), q')$, and $(q, quit, x, rew(x), q')$. The initial control is $q$ and stack is 0. Similarly for the model VPA. The state $q'$ is accepting.

It is easy to see that the counter machine has an infinite computation iff the configuration $q_0 0$ satisfies $\texttt{EG}^A\texttt{ff}$. □

**Pushdown Automata** For PDA we have the following, with undecidability for EF[DVPA].

| | DFA | NFA | DVPA |
|---|---|---|---|
| DFA/ NFA | EXPTIME-complete | 2EXPTIME-complete | undecidable |

**Theorem 6.16.** *Model checking PDA against* CTL[NFA,DFA] *is in EXPTIME, and for* CTL[NFA,NFA] *it is in 2EXPTIME.*

*Proof.* We show first that model checking CTL[NFA,DFA] over pushdown systems is in EXPTIME. The proof is via a game reduction that is a straightforward adaptation of the proof of Lemma 6.10. We make the following adjustments: since the formula automata are finite automata, we do not need a stack component for them. Hence, the stack is only updated and read by the PDA component. Because we no longer need to synchronise two stacks, the model can be a PDA.

Model checking CTL[NFA,NFA] over pushdown systems in 2EXPTIME follows from the previous part and the exponential cost of determinising an NFA. □

**Theorem 6.17.** *Model checking PDA against* EF[DVPA] *and* EG[DVPA] *are undecidable.*

We split the proof in two lemmas.

**Lemma 6.18.** *Model checking* EF[DVPA] *over pushdown systems is undecidable.*

*Proof.* The proof is almost identical to the case for VPA, except, since the model is a PDA, we no longer need to mark the output alphabet. More formally, our pushdown system has the following rules.

- $(q, o_2, x, push(1), q')$ where $x \in \{0, 1\}$ and $q \xrightarrow[inc_1]{o_2} q'$.
- $(q, o_2, 1, pop, q')$ where $q \xrightarrow[dec_1]{o_2} q'$.
- $(q, o_2, 0, rew(0), q')$ where $q \xrightarrow[zero_1^?]{o_2} q'$.
- $(q_f, fin, x, rew(x), q_{done})$ where $x \in \{0, 1\}$.

And $A$ has the following rules where $x$ ranges over $\{0, 1\}$: $(q, inc_2, x, push(1), q)$, $(q, dec_2, 1, pop, q)$, $(q, zero_2^?, 0, rew(0), q)$, and $(q, fin, x, rew(x), q')$. The initial control is $q$ and stack is 0. Similarly for the model VPA. The state $q'$ is accepting.

It is easy to see that the counter machine can reach $q_f$ iff the configuration $q_0 0$ satisfies $\mathtt{EF}^A\mathtt{tt}$. □

**Lemma 6.19.** *Model checking* EG[DVPA] *over pushdown systems is undecidable.*

*Proof.* Again, we proceed as in the VPA case. Our pushdown system has the following rules.

– $(q, o_2, x, push(1), q')$ where $x \in \{0, 1\}$ and $q \xrightarrow[inc_1]{o_2} q'$.
– $(q, o_2, 1, pop, q')$ where $q \xrightarrow[dec_1]{o_2} q'$.
– $(q, o_2, 0, rew(0), q')$ where $q \xrightarrow[zero_1^?]{o_2} q'$.
– $(q, quit, x, rew(x), q_{quit})$ where $x \in \{0, 1\}$ and $q$ is a state of the counter machine, or $q_{quit}$.

And $A$ has the following rules where $x$ ranges over $\{0, 1\}$: $(q, inc_2, x, push(1), q)$, $(q, dec_2, 1, pop, q)$, $(q, dec_2, 0, pop, q')$, $(q, zero_2^?, 0, rew(0), q)$, $(q, zero_2^?, 1, rew(1), q')$, and $(q, quit, x, rew(x), q')$. The initial control is $q$ and stack is 0. Similarly for the model VPA. The state $q'$ is accepting.

It is easy to see that the counter machine has an infinite computation iff the configuration $q_0 0$ satisfies $\mathtt{EG}^A\mathtt{ff}$. □

## 7 Conclusion and Further Work

To the best of our knowledge, this is the first work considering a parametric extension of CTL by arbitrary classes of formal languages characterising the complexities of satisfiability and model checking as well as the expressive power and model-theoretic properties of the resulting logics in accordance to the classes of languages. The results show that some of the logics, in particular CTL[VPL] may be useful in program verification because of the combination of an intuitive syntax with reasonably low complexities of the corresponding decision problems.

Some questions still remain to be answered, in particular the relationship between CTL[CFL] and $\Delta$PDL$^?$[CFL] and whether or not CTL[DCFL] $\lneq$ CTL[CFL] holds.

Furthermore, there are obvious directions for further work. It is possible to consider CTL$^*$ or CTL$^+$ as the base for similar extensions. It is also possible to extend such logics with automata on infinite words, for instance in the form of path quantifier relativization. This may be even more suitable in the framework of abstraction and refinement as mentioned in the introduction.